\documentclass[twocolumn]{aastex62}

\usepackage{xspace}
\usepackage{multirow}
\usepackage{booktabs}
\usepackage{subfigure}

\newcommand{\nuTXS}{IceCube-170922A\xspace}
\newcommand\TXS{TXS\,0506+056\xspace}
\newcommand{\nuGB}{IceCube-141209A\xspace}
\newcommand\FGL{3FGL\,J1040.4+0615\xspace} 
\newcommand\GB{GB6\,J1040+0617\xspace}  
\newcommand\PKS{PKS\,0502+049\xspace}

\newcommand\CCCC{4C+06.41\xspace}
\newcommand\SDSS{SDSS\,J104039.54+061521.5\xspace}

\newcommand{\Fermi}{\textit{Fermi}\xspace}

\graphicspath{{./}{figures/}}

\revised{June 6, 2019}
\accepted{June 17, 2019}
\published{July 31, 2019}

\shorttitle{\TXS and \GB}
\shortauthors{S. Garrappa et al.}

\begin{document}

\title{Investigation of two \Fermi-LAT gamma-ray blazars coincident with high-energy neutrinos detected by IceCube}


\author{S. Garrappa}
\affiliation{DESY, D-15738 Zeuthen, Germany}
\author{S. Buson}
\affiliation{University of W\"{u}rzburg, 97074 W\"{u}rzburg, Germany}
\affiliation{University of Maryland, Baltimore County, MD, USA}
\author{A. Franckowiak}
\affiliation{DESY, D-15738 Zeuthen, Germany}
\collaboration{\textit{Fermi}-LAT collaboration}
\noaffiliation
\nocollaboration
\author{B. J. Shappee}
\affiliation{Institute for Astronomy, University of Hawai'i, Honolulu, HI 96822, USA}
\author{J. F. Beacom}
\affiliation{Dept. of Astronomy, Ohio State University, Columbus, OH 43210, USA}
\affiliation{Dept. of Physics and Center for Cosmology and Astro-Particle Physics, Ohio State University, Columbus, OH 43210, USA}
\affiliation{Center for Cosmology and Astroparticle Physics, The Ohio State University, Columbus, OH 43210, USA}
\author{S. Dong}
\affiliation{Kavli Institute for Astronomy and Astrophysics, Peking University, 5 Yiheyuanlu, Haidian District Beijing, China 100871}
\author{T. W.-S. Holoien}
\affiliation{The Observatories of the Carnegie Institution for Science, 813 Santa Barbara St., Pasadena, CA 91101, USA}
\author{C. S. Kochanek}
\affiliation{Dept. of Astronomy, Ohio State University, Columbus, OH 43210, USA}
\affiliation{Center for Cosmology and Astroparticle Physics, The Ohio State University, Columbus, OH 43210, USA}
\author{J. L. Prieto}
\affiliation{N\'ucleo de Astronom\'ia de la Facultad de Ingenier\'ia y Ciencias, Universidad Diego Portales, Av. Ej\'ercito 441, Santiago, Chile}
\affiliation{Millennium Institute of Astrophysics, Santiago, Chile}
\author{K. Z. Stanek}
\affiliation{Dept. of Astronomy, Ohio State University, Columbus, OH 43210, USA}
\author{ T. A. Thompson}
\affiliation{Dept. of Astronomy, Ohio State University, Columbus, OH 43210, USA}
\collaboration{ASAS-SN collaboration}
\nocollaboration
\author{M. G. Aartsen}
\affiliation{Dept. of Physics and Astronomy, University of Canterbury, Private Bag 4800, Christchurch, New Zealand}
\author{M. Ackermann}
\affiliation{DESY, D-15738 Zeuthen, Germany}
\author{J. Adams}
\affiliation{Dept. of Physics and Astronomy, University of Canterbury, Private Bag 4800, Christchurch, New Zealand}
\author{J. A. Aguilar}
\affiliation{Universit{\'e} Libre de Bruxelles, Science Faculty CP230, B-1050 Brussels, Belgium}
\author{M. Ahlers}
\affiliation{Niels Bohr Institute, University of Copenhagen, DK-2100 Copenhagen, Denmark}
\author{M. Ahrens}
\affiliation{Oskar Klein Centre and Dept. of Physics, Stockholm University, SE-10691 Stockholm, Sweden}
\author{C. Alispach}
\affiliation{D{\'e}partement de physique nucl{\'e}aire et corpusculaire, Universit{\'e} de Gen{\`e}ve, CH-1211 Gen{\`e}ve, Switzerland}
\author{K. Andeen}
\affiliation{Department of Physics, Marquette University, Milwaukee, WI, 53201, USA}
\author{T. Anderson}
\affiliation{Dept. of Physics, Pennsylvania State University, University Park, PA 16802, USA}
\author{I. Ansseau}
\affiliation{Universit{\'e} Libre de Bruxelles, Science Faculty CP230, B-1050 Brussels, Belgium}
% ...

\collaboration{IceCube Collaboration}
\noaffiliation

\collaboration{}
\collaboration{}

\date{\today}

 
\begin{abstract}

After the identification of the gamma-ray blazar \TXS as the first compelling IceCube neutrino source candidate, we perform a systematic analysis of all high-energy neutrino events satisfying the IceCube realtime trigger criteria.  We find one additional known gamma-ray source, the blazar \GB, in spatial coincidence with a neutrino in this sample.  The chance probability of this coincidence is 30\% after trials correction. For the first time, we present a systematic study of the gamma-ray flux, spectral and optical variability and multi-wavelength behavior of \GB and compare it to \TXS. We find that \TXS shows strong flux variability in the \Fermi-LAT gamma-ray band, being in an active state around the arrival of \nuTXS, but in a low state during the archival IceCube neutrino flare in 2014/15. In both cases the spectral shape is statistically compatible ($\leq 2\sigma$) with the average spectrum showing no indication of a significant relative increase of a high-energy component. While the association of \GB with the neutrino is consistent with background expectations, the source appears to be a plausible neutrino source candidate based on its energetics and multi-wavelength features, namely a bright optical flare and modestly increased gamma-ray activity. Finding one or two neutrinos originating from gamma-ray blazars in the given sample of high-energy neutrinos is consistent with previously derived limits of neutrino emission from gamma-ray blazars, indicating the sources of the majority of cosmic high-energy neutrinos remain unknown.

\end{abstract}


\section{Introduction} \label{sec:intro}
\email{simone.garrappa@desy.de,\,sara.buson@gmail.com,\\ \,anna.franckowiak@desy.de,\,analysis@icecube.wisc.edu}
The IceCube Neutrino Observatory has detected a diffuse flux of high-energy neutrinos in the energy range from 30\,TeV to 2\,PeV \citep{Aartsen:2013jdh,PhysRevD.91.022001,Aartsen:2016xlq}. However, until recently no compelling evidence for spatial or temporal clustering of events had been identified and the origin of the neutrinos was unknown~\citep{Aartsen:2016oji,2015ApJ...807...46A}. The arrival directions of IceCube neutrinos are compatible with an isotropic distribution, suggesting a predominantly extra-galactic origin for the cosmic neutrinos.
Among the most promising source candidates are (low luminosity) gamma-ray bursts, choked-jet and interacting supernovae, tidal disruption events, star forming galaxies and active galactic nuclei (AGN) - see \citet{2015RPPh...78l6901A} for a recent review. In general, high-energy neutrinos are produced through interactions of cosmic rays with ambient matter or photon fields. Charged and neutral pions produced in those interactions produce neutrinos and gamma rays, respectively, in their decay chain.

Blazars, those AGN with a relativistic jet of plasma pointing towards the observer, have been suggested as high-energy cosmic-ray accelerators and, in turn, neutrino sources \citep[e.g.][]{mannheim89,1991PhRvL..66.2697S,protheroe92,mannheim92,1993A&A...269...67M,szabo94,1995APh.....3..295M,mastichiadis96,protheroe99,bednarek99,mucke01,mannheim01,atoyan01,protheroe03,atoyan03,mucke03,reimer04,dermer09,dermer12,dimitrakoudis12,boettcher13,halzen13,2014MNRAS.443..474P,2016NatPh..12..807K}.
The spectral energy distribution (SED) of blazars exhibits two broad bumps. While the lower-energy one likely arises from synchrotron radiation of primary electrons, the origin of the higher-energy one is still a matter of debate. In leptonic models it is described by inverse Compton scattering, while in hadronic models the decay of $\pi_0$ produced in p-$\gamma$ interactions can be responsible for the second bump. Both leptonic and hadronic models are capable of adequately reproducing the observed emission for most sources \citep{2013ApJ...768...54B}. Only hadronic models predict emission of high-energy neutrinos, which originate in the interaction of protons with lower-energy photons. Those target photons could be produced in external fields of the broad line region
(as suggested for flat spectrum radio quasars, FSRQs \citep[]{Dermer:2012rg,Diltz:2015kha,Petropoulou:2015swa}), the accretion disk (\citep[]{Kachelriess:2008qx,Dermer:2008cy,Atoyan:2008uy,Fujita:2015xva}) or synchrotron photons in the jet (e.g.~in BL Lacs \citep{Cerruti:2014iwa}). The production of $\mathcal{O}$(100\,TeV) neutrinos would be accompanied by $\mathcal{O}$(200\,TeV) gamma rays, implying a correlation between gamma-ray and neutrino fluxes at the source \citep{Ahlers:2018fkn}. However, those high-energy photons interact quickly in the source or during propagation and cascade down to lower energies. Furthermore, gamma rays produced in alternative processes such as bremsstrahlung and inverse Compton scattering could alter the neutrino to gamma-ray connection.

Hints of correlations between neutrinos and blazars have been suggested by several groups \citep[e.g.~][]{2016NatPh..12..807K,2016MNRAS.457.3582P,2018A&A...620A.174K,2019ApJ...870..136L}. It is evident that multi-messenger studies are crucial to probe various source classes as potential neutrino emitters~\citep{10.1088/978-0-7503-1369-8}, as well as shed light onto the emission mechanisms of blazars.

To enable an efficient search for electromagnetic counterparts to the high-energy astrophysical neutrino signal, IceCube has implemented a realtime program \citep{2017APh....92...30A}. The program selects high-energy neutrinos ($\gtrsim 60$\,TeV) of likely cosmic origin within seconds of their detection at the South Pole, and distributes the information of the reconstructed neutrino direction to a network of follow-up instruments. 
On September 22, 2017, the program released an alert reporting an event with an estimated neutrino energy of $>$100\,TeV and good angular reconstruction, \nuTXS. Shortly after, the \Fermi Large Area Telescope (LAT) collaboration reported the detection of a potential electromagnetic counterpart in spatial coincidence with this high-energy neutrino event \citep{Tanaka17}. The gamma-ray signal was consistent with the known gamma-ray blazar, \TXS, which at the time of the IceCube trigger was in a state of enhanced activity~\citep{Tanaka17,MWScience}. Subsequently, $>$100\,GeV gamma-ray emission was detected from \TXS for the first time by the Major Atmospheric Gamma Imaging Cherenkov Telescopes (MAGIC, \citealt{MWScience,Ahnen:2018mvi}), which was later confirmed by the Very Energetic Radiation Imaging Telescope Array System \citep[VERITAS,][]{2018ApJ...861L..20A}.
Searches with the ANTARES neutrino telescope yielded no convincing evidence of additional neutrino emission related to the source \citep{2018ApJ...863L..30A}, which is consistent with expectations.

Chance spatial coincidence between the neutrino and the flaring blazar was disfavored with $3\sigma$ significance~\citep{MWScience}. An archival search for additional $>$100\,GeV neutrinos from the location of \TXS led to the discovery of a candidate neutrino flare between September 2014 and March 2015 at $3.5\sigma$ significance~\citep{ICScience}. While the $\sim$290 TeV neutrino in 2017 was accompanied by increased activity in gamma rays indicating a neutrino-gamma-ray connection, the source was in a low gamma-ray state during the 2014/15 neutrino flare.

The possible detection of neutrino emission from the blazar has motivated several attempts to model the multi-wavelength SED of \TXS, assuming simultaneous leptonic and hadronic emission in so-called hybrid models (see e.g.~\citealt{2003ApJ...586...79A,2005ApJ...621..176B,2015A&A...573A...7W,Gao:2018mnu,Ahnen:2018mvi,Cerruti:2018tmc,Keivani:2018rnh}).

A second spatial coincidence was pointed out in ~\citet{MWScience} of an archival high-energy neutrino event with the \Fermi-LAT source \FGL.

In this paper we carry out a study of potential \Fermi-LAT gamma-ray counterparts to the high-energy events observed by IceCube. We present a detailed investigation of the candidate electromagnetic counterparts found in spatial connection to two high-energy IceCube neutrino alerts.

\section{Search for high-energy neutrinos in coincidence with gamma-ray sources}
The IceCube neutrino observatory is a cubic kilometer scale Cherenkov detector located at a depth of 1450\,m to 2450\,m in the clear ice of the geographic South Pole. A total of 5160 digital optical modules are located on 86 strings arranged in a hexagonal grid to detect Cherenkov light emitted by secondary charged particles produced in neutrino interactions in or close to the instrumented detector volume~\citep{2017JInst..12P3012A}.

The sample of neutrinos considered in this study is based on the high-energy neutrino events observed by the IceCube detector from 2010 to 2017, and satisfying the IceCube realtime trigger criteria.
This includes ten published realtime alerts (up to and including \nuTXS) and forty archival events\footnote{\url{https://icecube.wisc.edu/science/data/TXS0506_alerts}}. Among the latter, five are flagged because of their poor angular reconstruction, which would have caused them to be retracted as realtime alerts.
To reduce the number of chance coincidences, we apply the same sample selection cut of \citet{MWScience} and restrict the study to events with a 90\% angular uncertainty\footnote{We note that the uncertainty contours considered here are a result of the processing techniques applied at the time and may experience small changes with future analysis improvements that reflect more accurate treatment of the systematic uncertainties.} smaller than 5 deg$^2$. Eight events do not satisfy this criteria and are thus discarded. The final neutrino sample consists of thirty seven well-reconstructed events. Each event is cross checked with the Third \Fermi-LAT Point Source Catalog ~\citep[3FGL,][]{2015ApJS..218...23A} and the Third \Fermi-LAT Catalog of High-Energy Sources~\citep[3FHL,][]{2017ApJS..232...18A} to search for spatial coincidences with known gamma-ray sources.

Among the remaining 37 neutrino events, besides the \nuTXS/\TXS occurrence, one additional spatial coincidence with a gamma-ray source is confirmed in this search (see also ~\citet{MWScience}). The gamma-ray source \FGL \citep{2015ApJ...810...14A}, which is associated with \GB, a BL Lac object with redshift 0.7351$\pm$0.0045 \citep{Maselli:2015oha,GB_redshift_old}. We note that the redshift measurement might be unreliable given that the automatic extraction was flagged by the SDSS pipeline, which indicates a poorly determined redshift. \cite{2009ApJS..180...67R} have reported a photometric redshift range of 2.210-2.950. However the spectrum in \citet{GB_redshift_old} does not show any indication of the Lyman-alpha forest, which makes it unlikely that the source is at redshift larger than 2. Also the colors of the source (u-g=0.71, g-r=0.48, r-i=0.51, i-z=0.4, from SDSS) indicate a redshift smaller than 2.7 according to the color-redshift relation of~\citet{2004ApJS..155..243W}.
\GB is located within the 90\% uncertainty of the well-reconstructed neutrino \nuGB. 

We focus on the potential astrophysical counterparts of these two high-energy neutrinos, and present a detailed investigation of the gamma-ray properties enabled by the continuous all-sky coverage of the \Fermi-LAT.

\section{\Fermi-LAT \MakeLowercase{data}}
The \Fermi-LAT is a pair-conversion telescope sensitive to gamma rays with energies from $20\,$MeV to greater than $300\,$GeV~\citep{2009ApJ...697.1071A}. It has a large field of view ($>$ 2sr) and scans the entire sky every three hours during standard operation, making it well suited to monitor variable gamma-ray sources on different timescales, from seconds to years.

In this study we use 9.6 years of Pass\,8 \Fermi-LAT data collected between 2008 August 4 and 2018 March 16 (MJD 54682-58193), selecting photons from the event class developed for point source analyses\footnote{\url{http://fermi.gsfc.nasa.gov/ssc/data/analysis/documentation/Pass8_usage.html}}. We perform a likelihood analysis\footnote{We use MINUIT as optimizer with 10$^{-3}$ tolerance}, binned in space and energy, using the standard \Fermi-LAT ScienceTools package version v11r5p3 available from the \textit{Fermi} Science Support Center\footnote{\url{http://fermi.gsfc.nasa.gov/ssc/data/analysis/}} (FSSC) and the \textsf{P8R2\_SOURCE\_V6} instrument response functions, together with the fermipy package v0.16.0 \citep{Wood:2017yyb}. We analyze data in the energy range from 100\,MeV to 1\,TeV binned into eight logarithmically-spaced energy intervals per decade. To minimize the contamination from gamma rays produced in the Earth's upper atmosphere, we apply an instrument zenith angle cut of $\theta<90\degr$. We use the standard data quality cuts \textit{($DATA\_QUAL>0$)$\&\&$($LAT\_CONFIG==1$)} and we remove time periods coinciding with solar flares and gamma-ray bursts detected by the LAT. The effect of energy dispersion is included in the fits performed with the \Fermi-LAT ScienceTools. 

In the analysis of \GB, an additional data cut is applied to remove the time periods when the Sun was located less than $15\degr$ from the source position. This additional cut is necessary because \GB lies very close to the ecliptic.

For each source, we select a $10\degr \times 10\degr$ region of interest (ROI) centered on the source position, binned in $0\fdg1$ size pixels. The binning is applied in celestial coordinates using a Hammer-Aitoff projection. The input model for the ROI includes all known gamma-ray sources from the 3FGL catalog in a region of $15\degr \times 15\degr$, slightly larger than the ROI, and the isotropic and Galactic diffuse gamma-ray emission models provided by the standard templates \textsf{iso\_P8R2\_SOURCE\_V6\_v06.txt} (extrapolated linearly in the logarithm up to 1\,TeV) and \textsf{gll\_iem\_v06.fits}\footnote{\url{https://fermi.gsfc.nasa.gov/ssc/data/access/lat/BackgroundModels.html}}.

Given the different and longer integration time of our analysis with respect to the 3FGL, we search for new gamma-ray sources that were too faint for being included in the 3FGL. New putative point sources are modeled with a single power-law spectrum, with the index fixed to 2 and the normalisation free to vary in the fit. The search procedure is iterated until no further significant (TS$_{det}>$25) excess is found. The new point sources significantly detected in the longer-integration time data set are accounted for by the final ROI model.

\begin{figure*}[h!]
\gridline{\fig{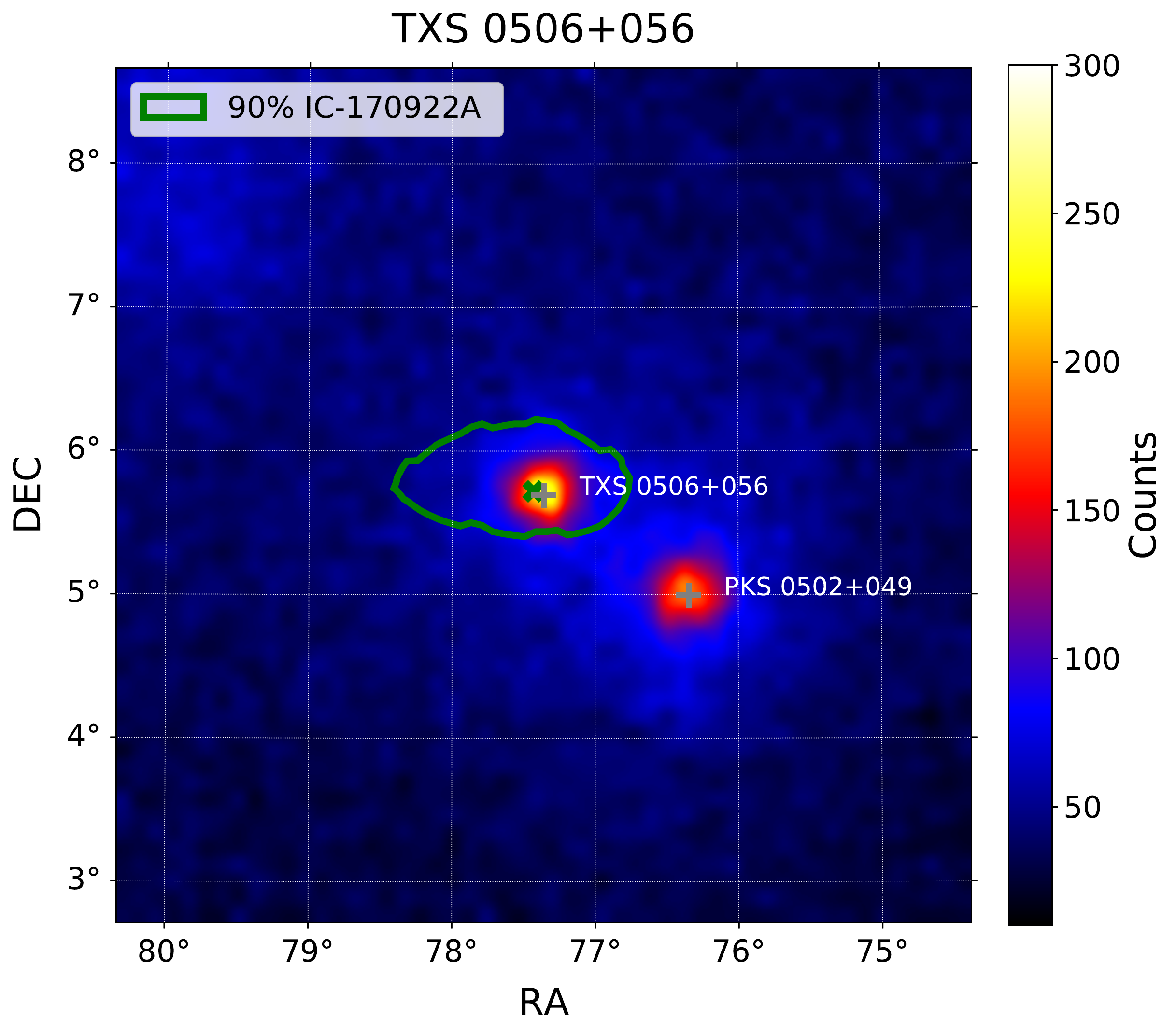}{0.5\textwidth}{(a)\,\TXS\,counts map.}\label{fig:txs0506_counts}
          \fig{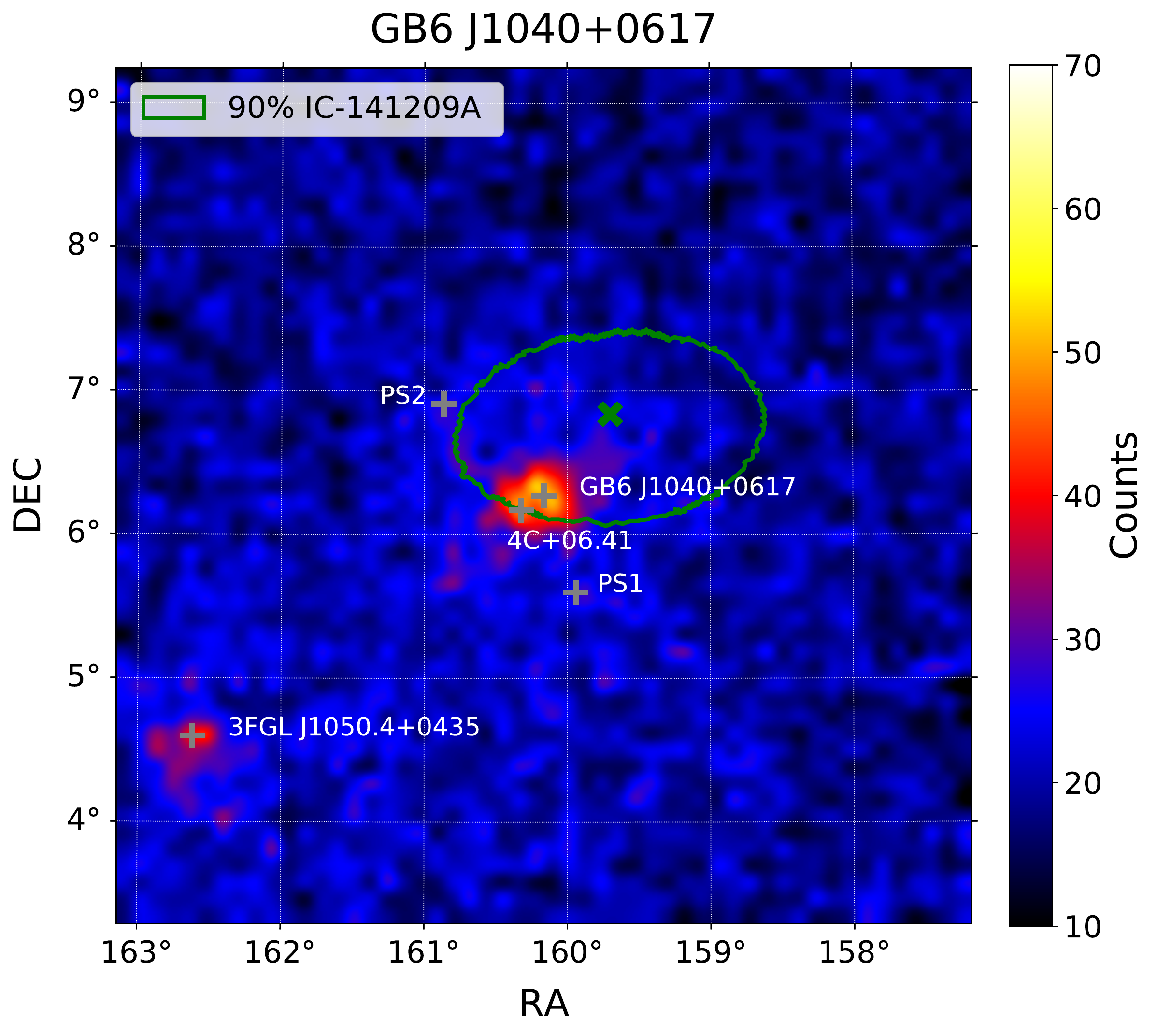}{0.5\textwidth}{(b)\,\GB\, counts map.}\label{fig:j1040_counts}
          }
\gridline{\fig{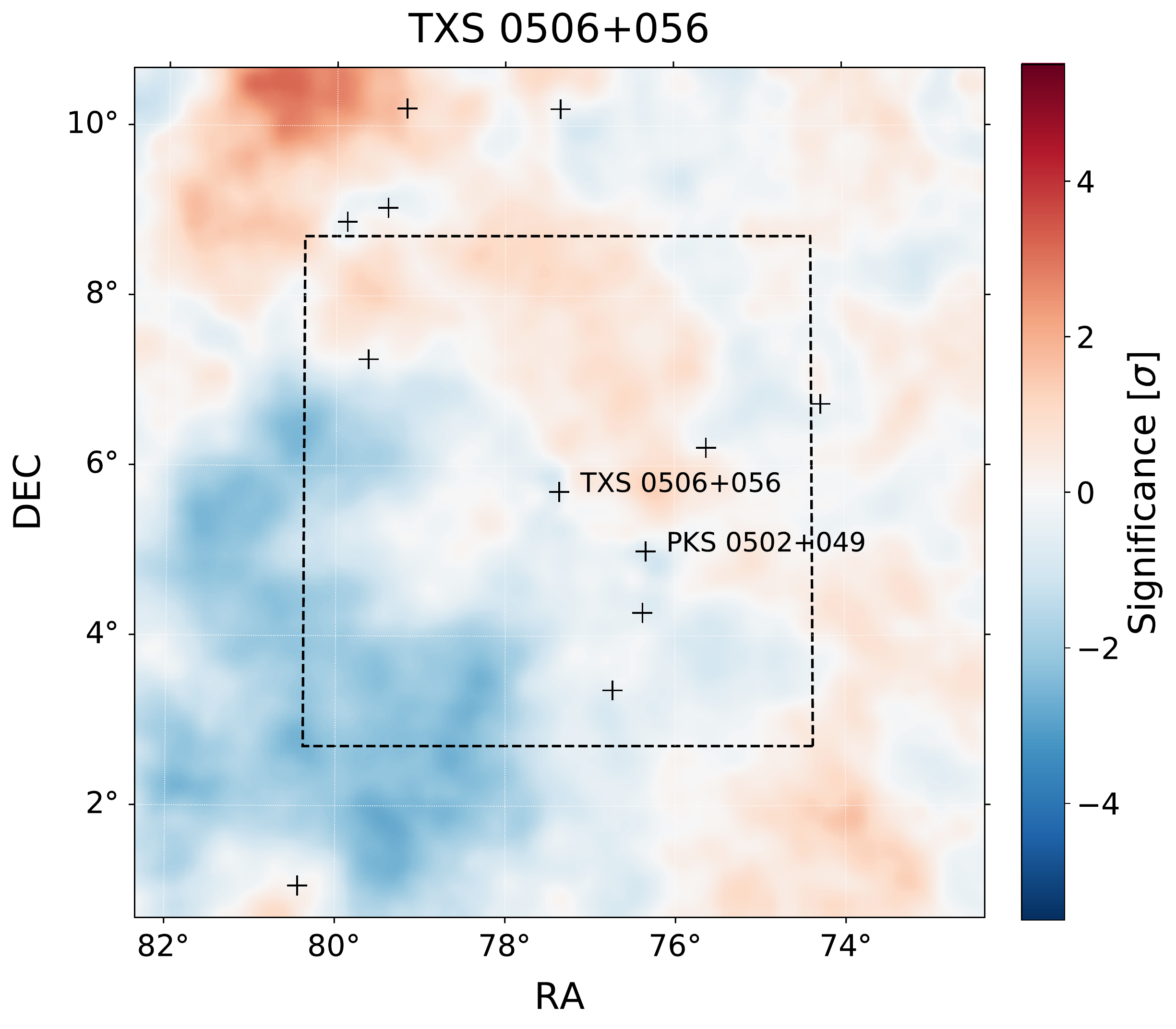}{0.5\textwidth}{(c)\,\TXS\,residual map.}\label{fig:txs0506_res}
          \fig{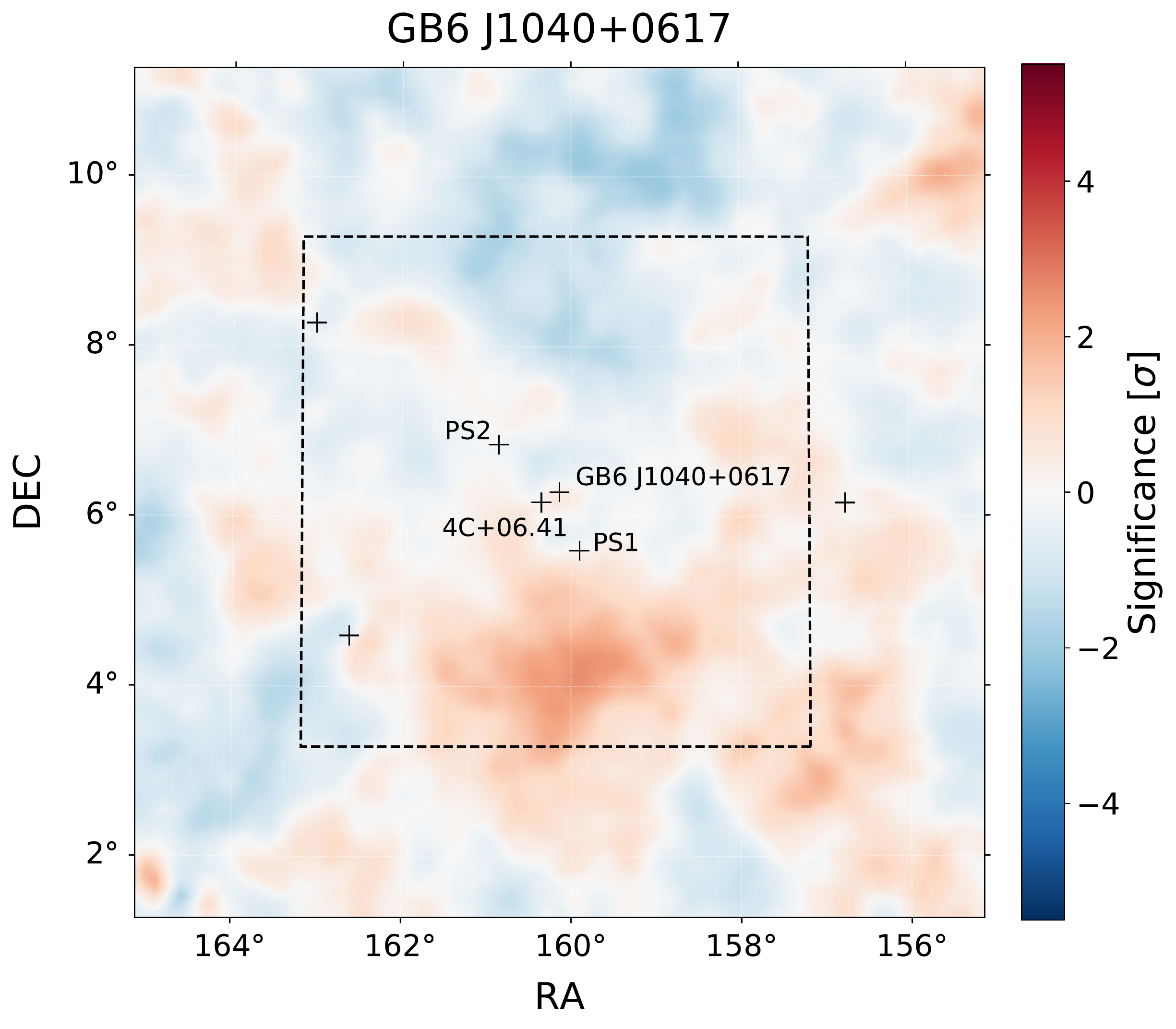}{0.5\textwidth}{(d)\,\GB\, residual map.}\label{fig:j1040_res}
          }

\caption{Top (bottom): counts maps (residual maps) $>$100\,MeV of the ROI centered on \TXS (left) and \GB (right). The 90\% neutrino angular uncertainty is shown as green contour in the counts maps, the best fit neutrino position is marked by a green "x". Sources modeled in the ROI are marked as black crosses in the residual maps. The residual maps show the entire ROI, while the counts maps show a zoom-in to the central region. The zoomed-in region displayed in the counts maps is indicated as a dashed black line in the residual maps. The maps are smoothed using a Gaussian with standard deviation of 0.1$\degr$. Note that the color scales for \TXS and \GB in the top panel are different.}
\label{fig:counts}
\end{figure*}

\section{\nuTXS}\label{sec:txs}
On 2017 September 22 at 20:54:30.43 UTC (MJD 58018.87) IceCube detected an extremely high-energy (EHE) through-going muon track event\footnote{More detail on the event selection can be found in \citet{2017APh....92...30A} and \url{https://gcn.gsfc.nasa.gov/doc/AMON_IceCube_EHE_alerts_Oct31_2016.pdf}}, \nuTXS, with a reconstructed direction of declination (Dec) $5\fdg72^{+0.50}_{-0.30}$ and right ascension (RA) $77\fdg43^{+0.95}_{-0.65}$ (J2000 equinox). The traversing muon deposited an energy of $(23.7\pm2.8)$\,TeV in the detector. The primary neutrino energy was estimated to be $\sim 290$\,TeV with a 90\% confidence lower limit of 183\,TeV and the fraction of neutrino events with this energy and arrival direction in the EHE alert stream that have an astrophysical origin is 56.5\% \citep[see][for details]{MWScience}.

The gamma-ray blazar \TXS is positionally consistent with \nuTXS, and it was undergoing a prolonged enhanced emission state at the time of the neutrino detection. This motivated a further search for neutrino emission from the direction of \nuTXS, in the whole IceCube archival dataset. A time-dependent analysis of 9.5 years of archival IceCube data ~\citep{ICScience} revealed an excess of detected neutrinos from the direction of \TXS between 2014 September and 2015 March, with a post-trials significance of 3.5$\sigma$. This excess was found using time-windows of variable width, with both a box-shaped and a Gaussian kernel finding the same excess with comparable significance. The best-fit Gaussian is centered at MJD 57004 with a width corresponding to two times the standard deviation of 110\,days and the box function covers the 158-day time range between MJD 56928 and 57086. 
The fit based on the box function provides a straight-forward definition of the start and end times of the fitted neutrino emission. We note that, assuming that the signal is Gaussian, one can show analytically that the optimal box-shaped time window in terms of signal/sqrt(background) is 1.5 times the width of the Gaussian, which matches well with the length of the box time window that was found. In the following, we refer to this excess as the \textit{neutrino flare} and adopt the parameters of the box kernel.

The gamma-ray source \TXS at Dec=~$+5\fdg69$, RA=~$77\fdg36$ lies well within the 50\% neutrino position uncertainty region, at a distance of $0\fdg1$ from the best fit neutrino position (see the gamma-ray counts map in Fig.~\ref{fig:counts}). The source is listed in the 3FGL as well as 3FHL  as 3FGL\,J0509.4+0541 and 3FHL\,J0509.4+0541, respectively~\citep[][]{2015ApJS..218...23A,2017ApJS..232...18A}. The 3FGL catalog is based on gamma-ray data in the energy range from 100\,MeV to 300\,GeV, whereas the 3FHL catalog is focused on the energy range from 10\,GeV to 2\,TeV. We note that \TXS is also in the 2FHL~\citep{2FHL} catalog based on gamma-ray data from 50\,GeV to 2\,TeV, identifying it already as a potential target for very-high-energy gamma-ray emission. \TXS is among the brightest 4.4\% (5.9\%) sources in the 3FGL (3FHL) in terms of gamma-ray energy flux within the energy bounds of the corresponding catalog~\citep[see also][]{2018arXiv180704461P}. We consider the gamma-ray energy flux more likely to be correlated with the neutrino flux than the gamma-ray number flux. Gamma rays accompanying the neutrino production are likely to cascade down to lower energies, not conserving the number flux, but the energy flux. The redshift of \TXS was measured to be $z = 0.336$ by ~\cite{2014ApJ...780...73A} and later confirmed by ~\cite{2018ApJ...854L..32P} at $z = 0.3365\pm0.0010$.

\subsection{Spectral Analysis}

We analyze 9.6 years of \Fermi-LAT data in the \TXS ROI. 
The source-finding algorithm finds one additional source with $TS_{det}$\footnote{$TS_{det}$ describes the difference in the maximum $\log \mathcal{L}$ of an ROI model with and without the source.}$>25$ at a distance of $2\fdg37$ from \TXS. This source is also included in the preliminary 8-year source list, FL8Y, provided by the \Fermi-LAT collaboration\footnote{FL8Y preliminary source list \url{https://fermi.gsfc.nasa.gov/ssc/data/access/lat/fl8y/}} as FL8Y J0518.4+0715 with no association. This source is outside the neutrino position uncertainty region.

\begin{figure}
  \centering
  \includegraphics[width=0.5\textwidth]{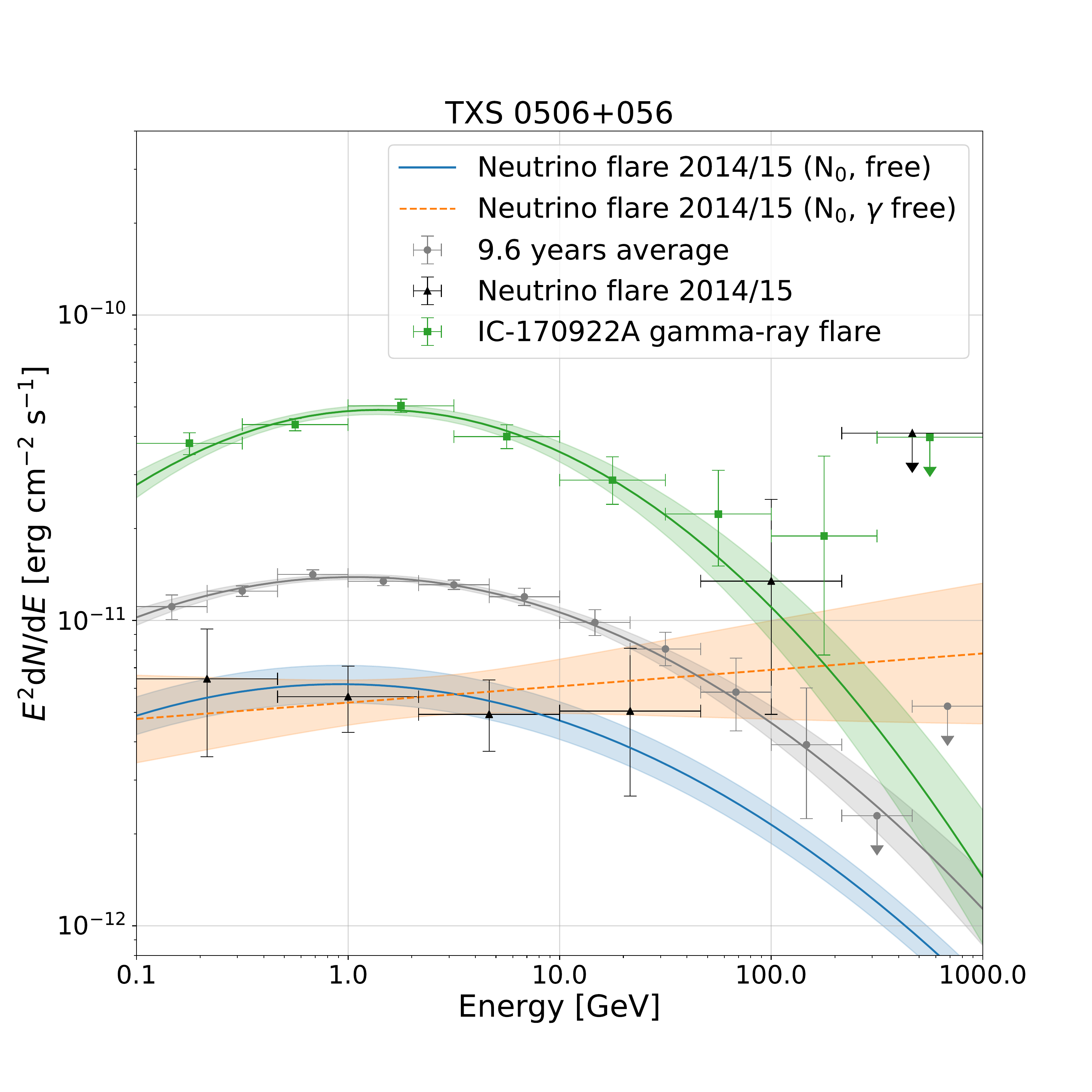}
  \caption{Gamma-ray spectrum of \TXS. \Fermi-LAT data for the whole 9.6-years time range are shown as gray crosses and the best-fit spectral model including statistical uncertainties is overlaid as a gray band. Arrows indicate 95\% upper limits. The spectrum of the 2017/18 gamma-ray flare is shown in green. The orange contour shows the spectrum during the 2014/15 neutrino flare modeled with a power-law function where both normalization and photon index are free to vary. The blue contour shows the log-parabola fit to the 2014/15 dataset, where only the normalization is left free to vary and the spectral parameters $\alpha$ and $\beta$ are fixed to the 9.6-years values. The spectral models reported should be considered reliable in the energy range where the source is significantly detected.}
  \label{fig:SED_txs}
\end{figure}

In the 3FGL catalog (based on 4 years of data) the gamma-ray spectrum of \TXS is modeled with a power-law function. An alternative
spectral model with an additional free parameter compared to a simple power-law is the log-parabola function:

\begin{equation}
\frac{dN}{dE} = N_0 \left(\frac{E}{E_b}\right)^{-(\alpha+\beta \log(E/E_b))}.
\end{equation}
where we use $E_{b} = 1.44$ GeV fixed during the whole analysis.
We find that for the almost ten-year data set a log-parabola model is preferred with a test statistic ($TS$) testing the different spectral shape models of $TS_{SS} = -2(\log \mathcal{L}_{PL} - \log \mathcal{L}_{LP}) = 374.3$ (i.e.~the log-parabola model describes the data better with a significance of $19\sigma$). Here $\mathcal{L}_{PL}$ and $\mathcal{L}_{LP}$ are the maximum likelihoods for the power-law and log-parabola spectral model respectively. We obtain a best-fit model of $\alpha = 2.03 \pm 0.02$, $\beta = 0.05 \pm 0.01$ and $N_{0} = (4.16\,\pm\, 0.08)\times 10^{-12}$\,cm$^{-2}$\,s$^{-1}$\,MeV$^{-1}$ (Fig.~\ref{fig:SED_txs}, gray spectrum).

The bright 3FGL source \PKS (Dec=~$+4\fdg99$, RA=~$76\fdg35$, J2000) is located $1\fdg23$ from \TXS. Previous studies discussed a possible source confusion between \PKS and \TXS~\citep{2018arXiv180704461P} and speculated if the archival neutrino flare originated in \PKS \citep{Liang:2018siw}.
The spectrum of \PKS is well-modeled by a log-parabola function with best fit values of $\alpha = 2.34 \pm 0.02$, $\beta = 0.10 \pm 0.01$ and $N_{0} = (1.08\,\pm\,0.02) \times 10^{-11}$\,cm$^{-2}$\,s$^{-1}$\,MeV$^{-1}$. Although \TXS is less bright than \PKS for energies below 1\,GeV, its energy flux integrated over the whole analysis energy range results in $(8.17 \pm 0.29)\times 10^{-11}$\,erg\,cm$^{-2}$\,s$^{-1}$ compared to $(6.70 \pm 0.13)\times 10^{-11}$\,erg\,cm$^{-2}$\,s$^{-1}$ for the nearby source.

The gamma-ray sky region is well described by the best-fit model, as can be seen in the residual map shown in Fig.~\ref{fig:counts} (bottom left), which does not show any significant structure.  

\subsection{Light Curve Analysis}
\label{sec:TXS_LC}

\begin{figure*}
  \centering
  
  \includegraphics[width=0.8\textwidth]{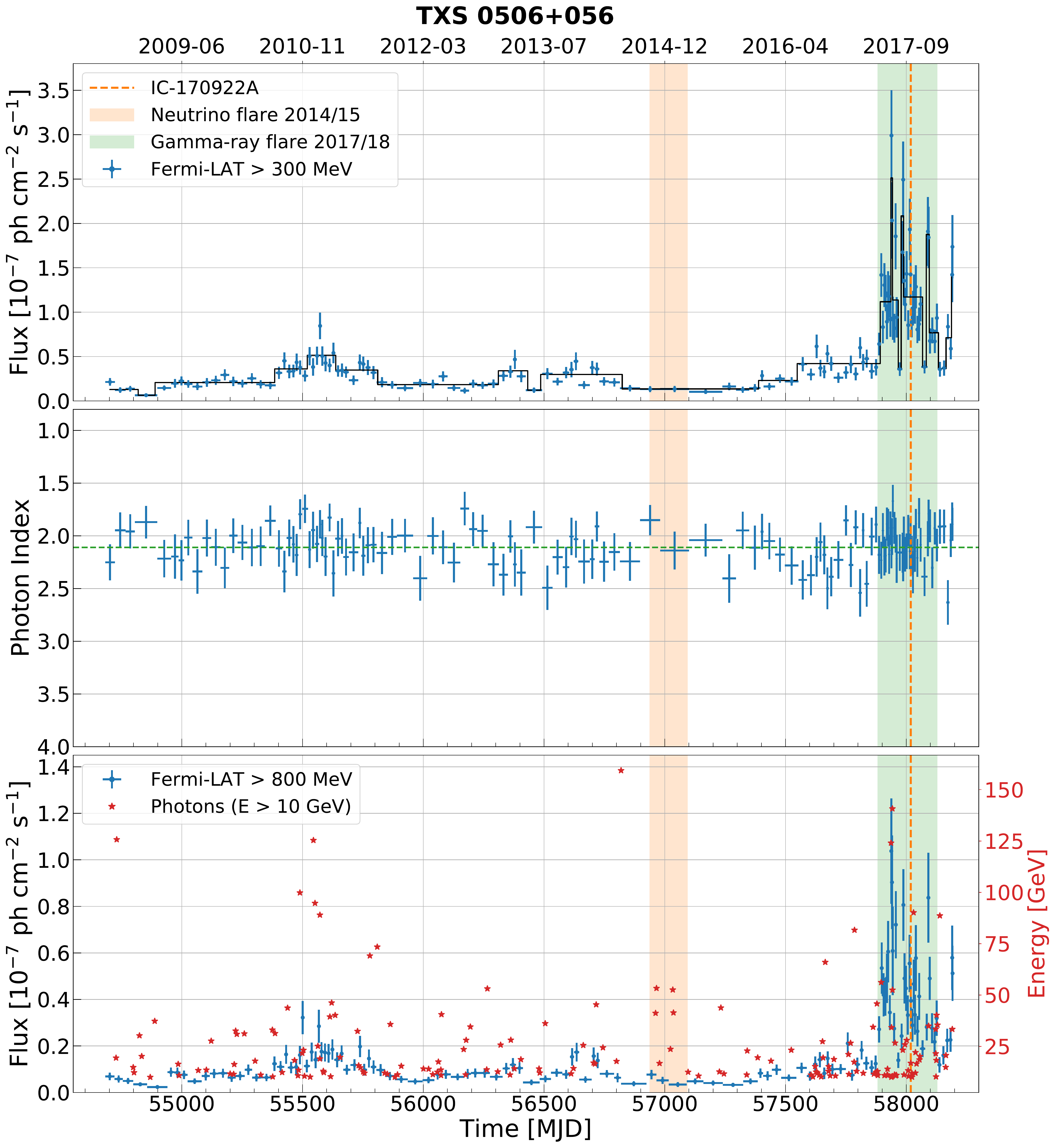}
  \caption{Adaptive binned light curve for \TXS. Panel 1 shows the gamma-ray flux integrated above 300\,MeV including the Bayesian Block representation shown in black, panel 2 shows the power-law spectral index and panel 3 shows the gamma-ray flux integrated above 800\,MeV. The average spectral index is shown as horizontal dashed green line in panel 2. The third panel additionally includes photons above 10\,GeV shown with red stars.}
  \label{fig:LC_TXS}
\end{figure*}

We produce an adaptively-binned (AB) light curve for \TXS, following the procedure in~\citet{2012A&A...544A...6L}. We chose a time binning which yields 15\% flux uncertainty in an energy range from 300\,MeV\footnote{The lower energy bound corresponds to the decorrelation energy, also referred to as optimum energy, defined in~\citet{2012A&A...544A...6L}.} to 1\,TeV, and perform a likelihood fit in each bin using a power-law model\footnote{On the short time scales considered here, the photon statistics are low and the source is well described by a power-law model.} for \TXS while allowing the spectral parameters of the closest neighboring sources to vary. The flux and spectral index variation are shown in Fig.~\ref{fig:LC_TXS}. 

To identify and characterize statistically-significant variations in the light curve, we apply the Bayesian Block algorithm outlined in \citet{2013ApJ...764..167S}, using the Astropy implementation\footnote{\url{http://docs.astropy.org/en/stable/api/astropy.stats.bayesian_blocks.html}}. To determine the optimal value of the prior for the number of blocks, we use the empirical relation in \citet{2013ApJ...764..167S} for the probability to falsely report a detection of a change point. This probability, which represents the relative frequency with which the algorithm reports the presence of a change point in data with no signal present, was set to 0.05.

While the largest historical gamma-ray outburst for \TXS occurred in 2017 in coincidence with \nuTXS (see Fig.~\ref{fig:LC_TXS}), the source does not display any remarkable activity during the neutrino flare in 2014/15. Also the gamma-ray spectral shape is compatible with the average over the whole mission. 
The spectral index shows small variations with respect to the average index of 2.11, and the source shows no obvious extended time periods of hardening or softening, over the full 9.6-year time range.

To further investigate the object's behavior during the neutrino flare, we derive the best-fit model for the region using a time window coincident with the 158-day neutrino excess. We then use the likelihood technique to robustly quantify any potential spectral change of the \TXS gamma-ray spectrum with respect to the average one. The likelihood ratio tests the hypothesis $H_0$, i.e.~the gamma-ray spectral shape is identical to the average one, against the hypothesis $H_1$, i.e.~an alternative spectral shape. The $H_0$ model allows only the normalization of \TXS to vary in the fit, while the spectral index is fixed to the average values obtained from the 9.6-years analysis. The $H_1$ has the spectral index of the power-law model for \TXS as an additional free parameter. All the other sources in the ROI, along with the Galactic and isotropic diffuse models, have the spectral parameters (including the normalisation) fixed to the 9.6-years fit results for both hypotheses. We define the test statistic to describe spectral change as $TS_{SC} = -2(\log \mathcal{L}_{0} - \log \mathcal{L}_{1})$, where $\mathcal{L}_{0}$ is the likelihood of the whole ROI for the null hypothesis, and $\mathcal{L}_{1}$ is the one corresponding to the alternative hypothesis $H_1$.

We repeat the analysis for various lower energy thresholds, $E_{\rm{min}}$, of 0.1, 0.5, 1, 2 and 10\,GeV and modeling \TXS with two different spectral shapes, i.e.~power-law and log-parabola. The results are summarized in Table~\ref{tab:table_lrt}. Note that in the case of the log-parabola spectral shape two additional parameters ($\alpha$ and $\beta$) are left free in the $H_1$ model with respect to $H_0$\footnote{The value of $E_{b}$ is always fixed, see~\citet{Massaro:2003sx}.}. According to Wilks' theorem~\citep{wilks1938} the $TS$ distribution can be assumed to follow a $\chi^2$ distribution with one or two degrees of freedom for the power-law or log-parabola spectral model, respectively~\citep{mattox96}. The p-value obtained from the $\chi^2$ distribution is converted to a Gaussian equivalent two-sided significance in units of sigma. For all tested cases, the p-value of the spectral change is $4\%$ or greater, providing no significant evidence in favor of a hardening or softening.

\citet{2018arXiv180704461P} found a spectral hardening during the neutrino flare in the energy range $>2$\,GeV with a $2\%$ p-value. 
In their analysis the lower threshold of $>2$\,GeV was chosen to avoid source confusion at lower energies with the neighboring source \PKS. By including the \PKS parameters as additional free parameters in our ROI model, we overcome the problem of source confusion, resulting in no significant residuals in the region of the two sources (see Fig.~\ref{fig:counts}). 

For the specific choice of $E_{\rm{min}}=2$\,GeV applied to a smaller time window of 110 days (corresponding to the $\pm1\sigma$ width of the Gaussian kernel of the neutrino flare search), we confirm the p-value of $2\%$ found by~\citet{2018arXiv180704461P} using a power-law model. 
In the box window width of 158 days we obtain a slightly lower significance of $2.1\sigma$ (p-value of $3.9\%$) for the same spectral model. For other choices of $E_{\rm{min}}$ we find lower significances (see Table \ref{tab:table_lrt}). 

\begin{table*}
\caption{Significance of spectral variations during the box time window of the neutrino flare}
\centering
\begin{tabular}{cccccccc}
\hline
\hline
$E_{\rm{min}}$ & \multicolumn{3}{c}{log parabola} & \multicolumn{3}{c}{power law} & \multirow{2}{*}{power law index}\\
$[$GeV$]$      & $TS_{SC}$ & $\sigma$\tablenotemark{a} & p-value & $TS_{SC}$ & $\sigma$\tablenotemark{a} & p-value \\ 
\hline
0.1  & 2.49 & 1.06 & 0.29 & 1.28 & 1.13 & 0.26 & $1.95 \pm 0.12$\\
0.5  & 4.13 & 1.53 & 0.13 & 3.87 & 1.97 & 0.05 & $1.88 \pm 0.13$\\
1.0  & 2.33 & 1.01 & 0.31 & 1.20 & 1.09 & 0.27 & $1.98 \pm 0.17$\\
2.0  & 5.12 & 1.77 & 0.08 & 4.25 & 2.06 & 0.04 & $1.76 \pm 0.20$\\
10.0  & 3.64 & 1.40 & 0.16 & 2.19 & 1.48 & 0.14 & $1.77 \pm 0.40$\\
\hline
\end{tabular}

\tablenotetext{a}{Significance in $\sigma$ assuming a Gaussian equivalent two-sided probability.}
\label{tab:table_lrt}
\end{table*}

In addition, we investigate possible patterns in the high-energy photons ($>$10\,GeV) which have a probability of association with \TXS of $>80\%$\footnote{This probability is obtained using the method \textsf{gtsrcprob} from the Fermi ScienceTools.} (see Table~\ref{tab:photons}). Under the hypothesis of a simple correlation between the gamma-ray and neutrino emission in blazars, the highest gamma-ray energies accessible by the LAT may be the best available tracer for high-energy neutrino emission in the absence of TeV gamma-ray observations. During the neutrino flare, we find six photons above 10\,GeV, among which are two with energies above 50\,GeV. For comparison, we look at the number of expected photons assuming the spectral shape during the non-flaring period of MJD 55800-56500\footnote{The non-flaring period is selected to start after the mild flaring period centered around 55500 and stops before a period of a moderate high-energy flaring activity seen in the lower panel of Fig.~\ref{fig:LC_TXS}.} and fitting the normalization in the 158-day time window, obtaining 4.44 (0.69) photons above 10 (50) GeV.
We find that the number of high-energy events observed during the neutrino flare is compatible with the typical gamma-ray behavior of \TXS during the 9.6 years of LAT monitoring.
The small excess of high-energy photons at face value has a p-value of 15\% corresponding to a one-sided Gaussian equivalent significance of $1\sigma$. The highest-energy photon associated with \TXS over the 9.6-year period was detected on MJD 56819 and has an energy of 159.3\,GeV.

\begin{table}

\caption{High-energy photons associated with \TXS with a probability of $>80\%$ detected during the neutrino flare time interval}
\centering
\begin{tabular}{cccc}
\hline
\hline
arrival time & dist.\tablenotemark{a} & energy & prob.\tablenotemark{b} \\
$[$MJD$]$ & $[\degr]$  & [GeV] & [\%] \\
\hline
56961.908 & 0.18 & 41.19 & 97.18 \\
56965.688 & 0.02 & 53.31 & 99.97 \\
56978.261 & 0.20 & 16.77 & 95.12 \\
57023.479 & 0.05 & 23.67 & 99.90 \\
57033.211 & 0.09 & 52.56 & 99.57 \\
57035.923 & 0.26 & 41.40 & 94.16 \\

\hline
\end{tabular}
\tablenotetext{a}{Angular distance to \TXS.}
\tablenotetext{b}{Probability to be associated with \TXS obtained with \textsf{gtsrcprob}.}
\label{tab:photons}
\end{table}

\begin{figure}
  \centering
  \includegraphics[width=0.45\textwidth]{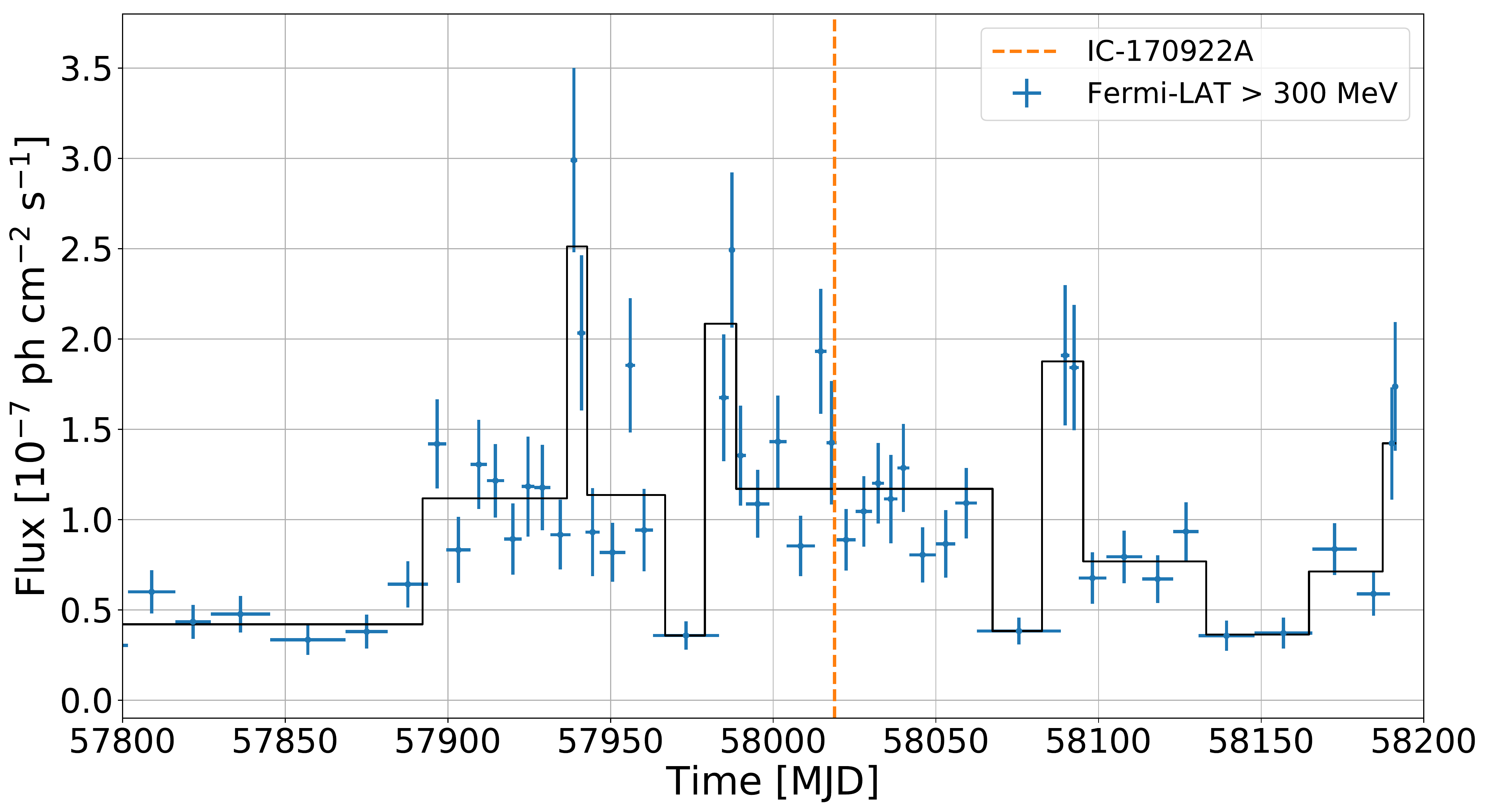}
  \caption{Zoomed-in gamma-ray light curve of \TXS around the arrival time of \nuTXS (shown in orange) and the bright gamma-ray flare. The black curve shows the result of the Bayesian Block algorithm.}
  \label{fig:LC_TXS_zoom}
\end{figure}

A closer investigation of the bright gamma-ray flare in 2017 shows significant structure, which is highlighted by the Bayesian Block algorithm (see Fig.~\ref{fig:LC_TXS_zoom}). 
We obtain gamma-ray spectra for the three brightest sub-flares ranging from MJD 57881-57963, 57983-58062 and 58088-58130 and repeat the search for spectral change applied during the neutrino flare period. We find $TS_{SC}$ values (starting from 100 MeV) of 3.5, 3.4 and 1.77 using a log-parabola function with two extra degrees of freedom pointing to similar spectral shapes compared to the average 9.6-years spectrum. The normalization during the sub-flares is 6.09, 6.37 and 5.1 times larger compared to the low-state defined over 700 days. Integrating over the whole flare duration we find 39 (5) photons above 10 GeV (50 GeV), which is compatible with the expected number of photons assuming the average spectral shape and a normalization fitted in the flare time window of 44.37 (4.16).

\section{\nuGB}
The High-Energy Starting (HESE) muon-track Event \nuGB \citep[event 63 in][]{2017arXiv171001191I} was detected on 2014 December 9 at 03:26:04.704 UTC (MJD 57000.14311). To obtain the reconstructed neutrino direction, a full likelihood scan is applied on a narrow grid with about $0\fdg06$ distance between the grid points. The resulting map of the likelihood landscape allows us to find the global minimum and the uncertainty contours at a given confidence level. The final best fit position and the 90\% confidence regions are shown in Fig.~\ref{fig:counts}. The position of the contour line is determined using a simulation of events with similar energy and trajectory through the detector as the event observed, while also varying the modeling of the optical properties of the deep glacial ice within the range allowed by systematic uncertainties, in order to obtain a conservative range. The minimum yields our best estimate of the event direction: Dec=~$6\fdg84$ and RA=~$159\fdg70$ (J2000 equinox) with a 90\% containment angular uncertainty region of $2.24$\,square degrees. We note that the best fit location moved and the 90\% uncertainty region increased with respect to the values published in \citet{MWScience} and the published event list\footnote{\url{https://icecube.wisc.edu/science/data/TXS0506_alerts}}. This is due to updated low-level re-calibrations and an event-by-event treatment of the systematic uncertainties, which are applied to events of special interest such as IceCube-170922A. The updated best-fit location remains within the original 50\% localization contour, and does not affect the conclusions in \citet{MWScience}. 

The event deposited an energy of $97.4^{+9.6}_{-9.6}$\,TeV in the detector. 
Following the procedure in \citet{2017APh....92...30A} we obtain a 29\% fraction of astrophysical signal events in the HESE alert sample for events with a similar or larger deposited charge, and which enter the detector from a similar arrival direction. Therefore, an atmospheric origin of the event cannot be excluded.

Within the 90\% uncertainty region of \nuGB, we identify only one cataloged gamma-ray source (among all 3FGL and 3FHL sources), \FGL. This source is located at a distance of $0\fdg70$ from the best-fit neutrino position. 

Following the approach presented in~\citet{MWScience}\footnote{Note that a simplified flat spatial probability density function is applied here in place of a Gaussian representation to accommodate computational constraints associated with the analysis of a large event sample.}, we estimate the p-value of the coincidence with \GB by considering the $N_s=2257$ extra-galactic \Fermi-LAT sources and their monthly light curves. Among all 30-day light curve bins of all sources, 9.5\% show a brighter gamma-ray energy flux in the energy range from 1--100\,GeV. The area of the 90\% neutrino position uncertainty region corresponds to $A_\nu=2.24$\,square degrees. The probability of finding an unassociated brighter source within the uncertainty region is hence $p=N_s A_\nu / (4\pi) \times 0.095 = 1$\%, which corresponds to a Gaussian equivalent, one-sided probability of $2.3\sigma$\footnote{Here we assume a uniform distribution of gamma-ray sources neglecting a reduced sensitivity for point source detection along the Galactic plane.}. After correcting for trials introduced by having searched for associations with each of the 37 well-reconstructed high-energy neutrino events in the sample, the final p-value is 30\%. In the following, we study if the multi-wavelength features of this source indicate a connection to the high-energy neutrino.

The source is included in 3FGL as well as 3FHL as 3FHL\,J1040.5+0618. We note that it is not included in 2FHL, and so it is not a $>50$\,GeV emitter. It is among the brightest 26.1\% (47.0\%) 3FGL (3FHL) sources in terms of gamma-ray energy flux for the 4-year (8-year) integration time. 
The most likely optical counterpart of this object is SDSS J104031.62+061721.7, located 1 arcmin from the 3FHL position, and associated with the low-synchrotron-peaked (LSP) BL Lac object \GB. Further analysis of the \nuGB region points out additional significant gamma-ray emission offset from the direction of \GB, and consistent with the blazar \CCCC. As discussed in the next sections, our detailed investigation indicates that this source awakened in gamma-rays in $\sim$mid-2015. We find no significant emission observed during the first 7 years of LAT monitoring, including the specific times around the \nuGB detection. During the 9.6 years considered in this work, the brightest persistent gamma-ray emission observed is consistent with the blazar \GB.

\subsection{Gamma-ray region of \nuGB}

We perform the same likelihood analysis of the ROI as described in Sec \ref{sec:txs}. Investigating the 9.6-years gamma-ray events in the vicinity of \FGL, we note significant gamma-ray emission offset from the sky direction of \GB, and positionally consistent with the radio position of the bright flat-spectrum radio quasar \CCCC, at redshift 1.27~\citep{2002MNRAS.329..700S}. This object is located at a distance of $0\fdg22$ from \GB and is neither in the 3FGL catalog nor the FL8Y list. In the second \Fermi-LAT source catalog of AGN \citep{2015ApJS..218...23A}, \CCCC was tentatively associated with the gamma-ray object 2FGL J1040.7+0614. 

In our ROI model, we therefore replace the single source \FGL with two point-like sources located at the radio positions of \GB and \CCCC ~\citep{1996ApJS..103..427G, 2009A&A...493..317L}. Integrating over the whole 9.6-year LAT dataset, \CCCC is detected with a $TS_{det}$ of 36, while \GB dominates the bulk of the observed gamma-ray emission with a $TS_{det}$ of 277. An examination of the temporal behavior of these objects in gamma rays is presented in section \ref{ssec:sec_lc_gb6} and helps to disentangle the gamma-ray emission observed from this region of the sky.

Furthermore, two additional new sources are found in the gamma-ray ROI, Fermi\,J1039.7+0535 and Fermi\,J1043.4+0654. This is not surprising given the longer integration time of our study with respect to the \Fermi-LAT catalog (more than double the 3FGL one). In the counts map in Fig.~\ref{fig:counts} we refer to these sources as PS1 and PS2, respectively. PS2 is also included in FL8Y as FL8Y J1043.4+0651 and associated with the BL Lac object 5BZB J1043+0653. PS1 and PS2 are dim sources with $TS_{det}$ values of 36.85 and 26.27 in the 9.6-year data set. They are adequately modeled by power-law spectra with best-fit spectral indices of $2.11\,\pm\,0.17$ and $1.80\,\pm\,0.21$, respectively. Both gamma-ray sources lie outside of the IceCube 90\% uncertainty contour. Based on the faintness of these sources and distance from the IceCube event, we do not investigate them further here. 

Investigating the gamma-ray region of \nuGB, we note two bright sources: 3FGL\,J1050.4+0435 detected with a $TS_{det}$ of 463, located $3\fdg0$ away from \GB; and 3FGL\,J1058.5+0133, detected with a $TS_{det}$ of 8619, located $6\fdg5$ from \GB. Since their gamma-ray flux is comparable to \GB, we let their spectral parameters free to vary in the likelihood fit.

\begin{figure*}[h!]
  \centering
  \includegraphics[width=0.8\textwidth]{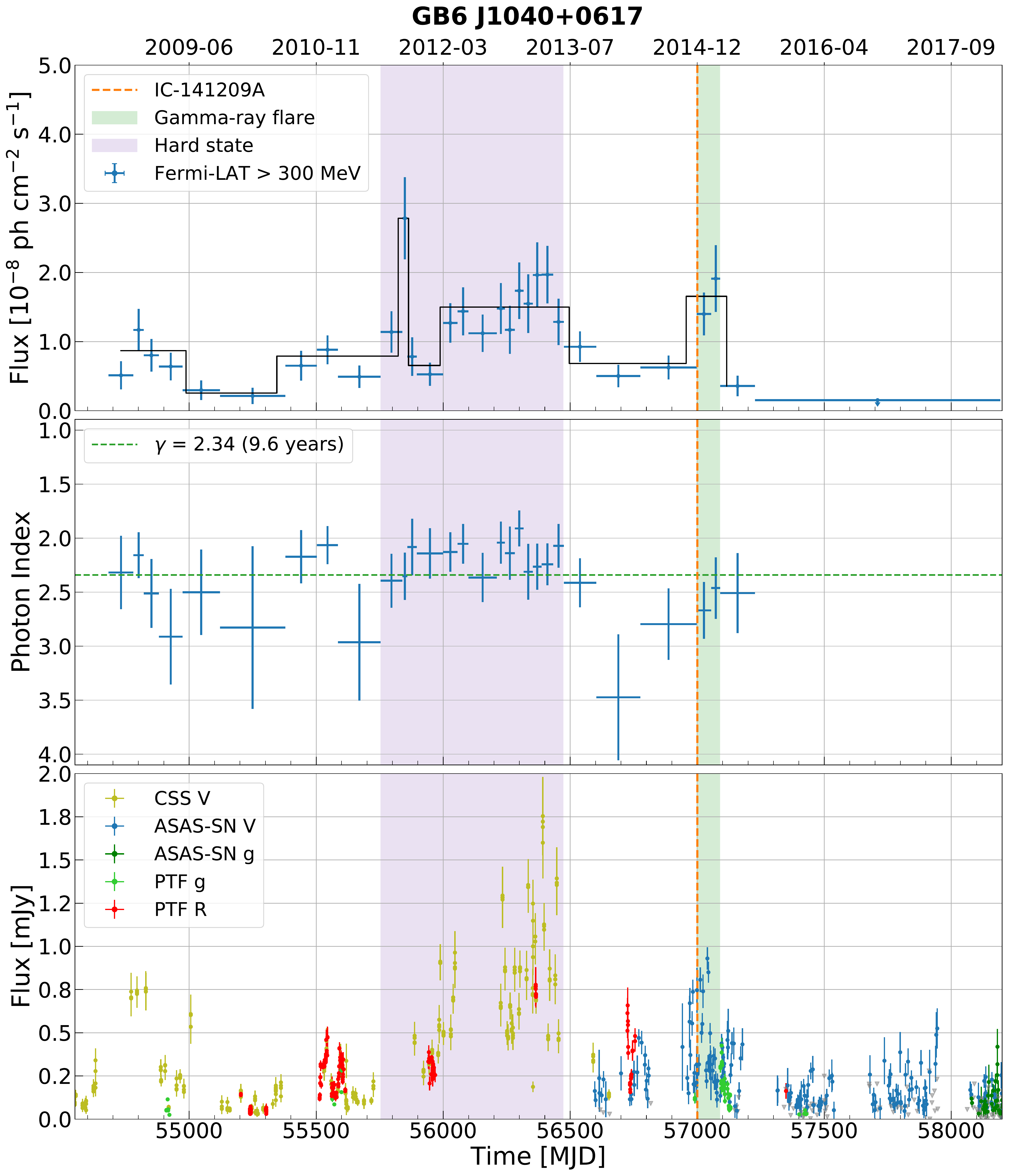}
  \caption{Adaptively-binned light curve for \GB. The first panel shows gamma-ray flux integrated above 300\,MeV including the Bayesian Block representation shown in black. The second panel shows the power-law spectral index. In the last time bin the source is not detected significantly, therefore a 95\% flux upper limit is shown in panel 1. In that case the spectral index cannot be fitted. The average spectral index is overlaid as a horizontal green dashed line. The third panel shows optical data obtained from ASAS-SN, PTF and CSS. ASAS-SN upper limits are displayed as gray triangles. The arrival of \nuGB is indicated as an orange dashed line. The purple shaded region marks the bright and hard gamma-ray state, while the green shaded region indicates the gamma-ray flare in coincidence with the neutrino arrival time.}
  \label{fig:LC_J1040}
\end{figure*}

\subsection{Gamma-ray light curve of \GB}\label{ssec:sec_lc_gb6}

The adaptively-binned light curve starting at the optimum energy of 300 MeV of \GB highlights several gamma-ray flux variations, observed throughout the 9.6 years (see Fig.~\ref{fig:LC_J1040}). The most clearly identified feature is a bright hard-spectrum state that lasts 721 days, from MJD 55753 to MJD 56474. During this period the source has a peak flux value of ($2.8\,\pm\,0.6\times 10^{-8}$)\,ph\,cm$^{-2}$\,s$^{-1}$ integrated in the energy range from 300 MeV to 1 TeV (a factor of 2.5 increase compared to the average flux) with an average power-law index of $2.08\,\pm\,0.04$ and an energy flux of ($2.84\,\pm\,0.94)\times 10^{-11}$\,erg\,cm$^{-2}$\,s$^{-1}$. The source shows increased activity that starts a few days before the \nuGB detection and lasts 93 days, from MJD 56997 to 57090. The duration of this flare is defined by the bin edges of the two high-flux adaptively-binned time bins.

\begin{figure*}
\gridline{\fig{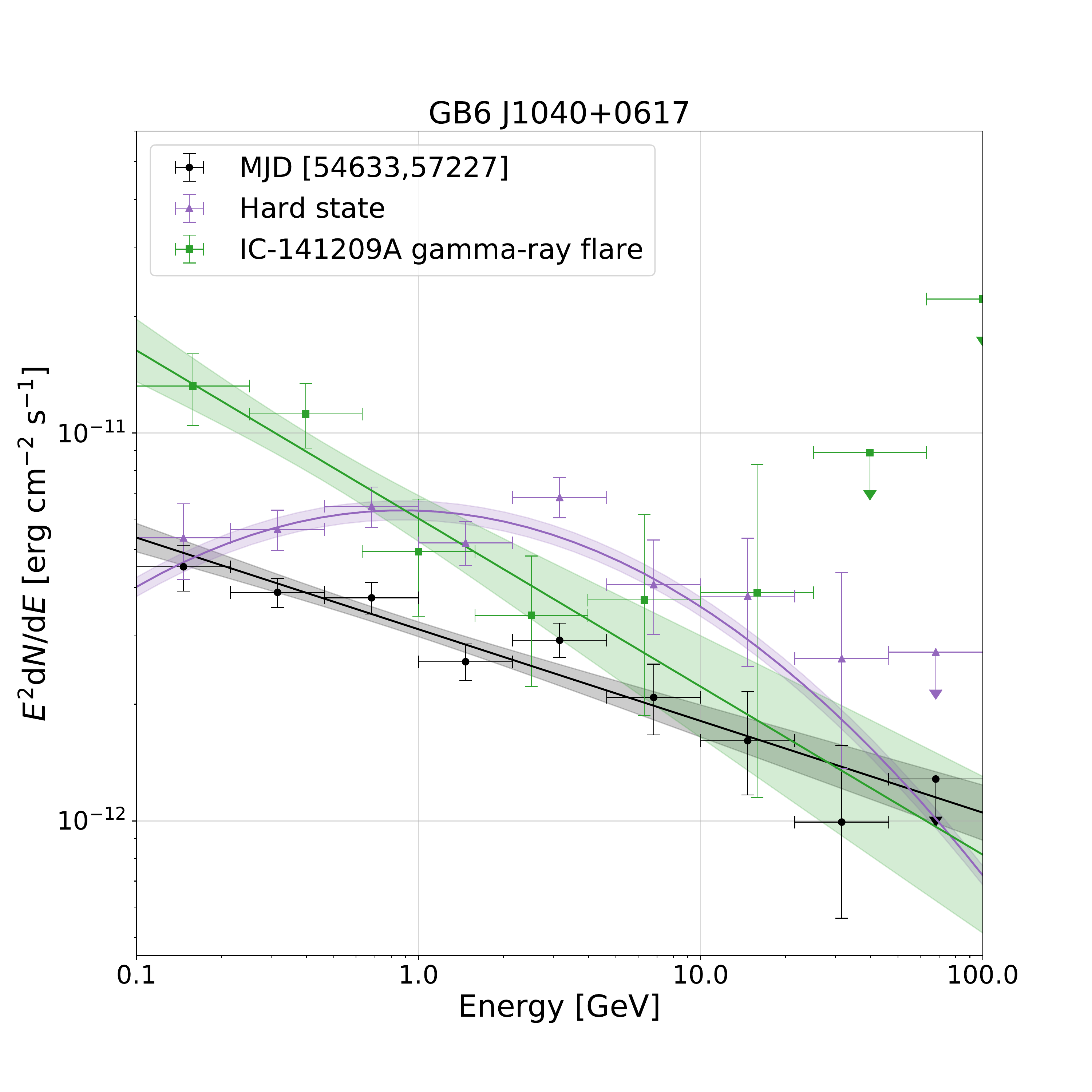}{0.5\textwidth}{(a)}
          \fig{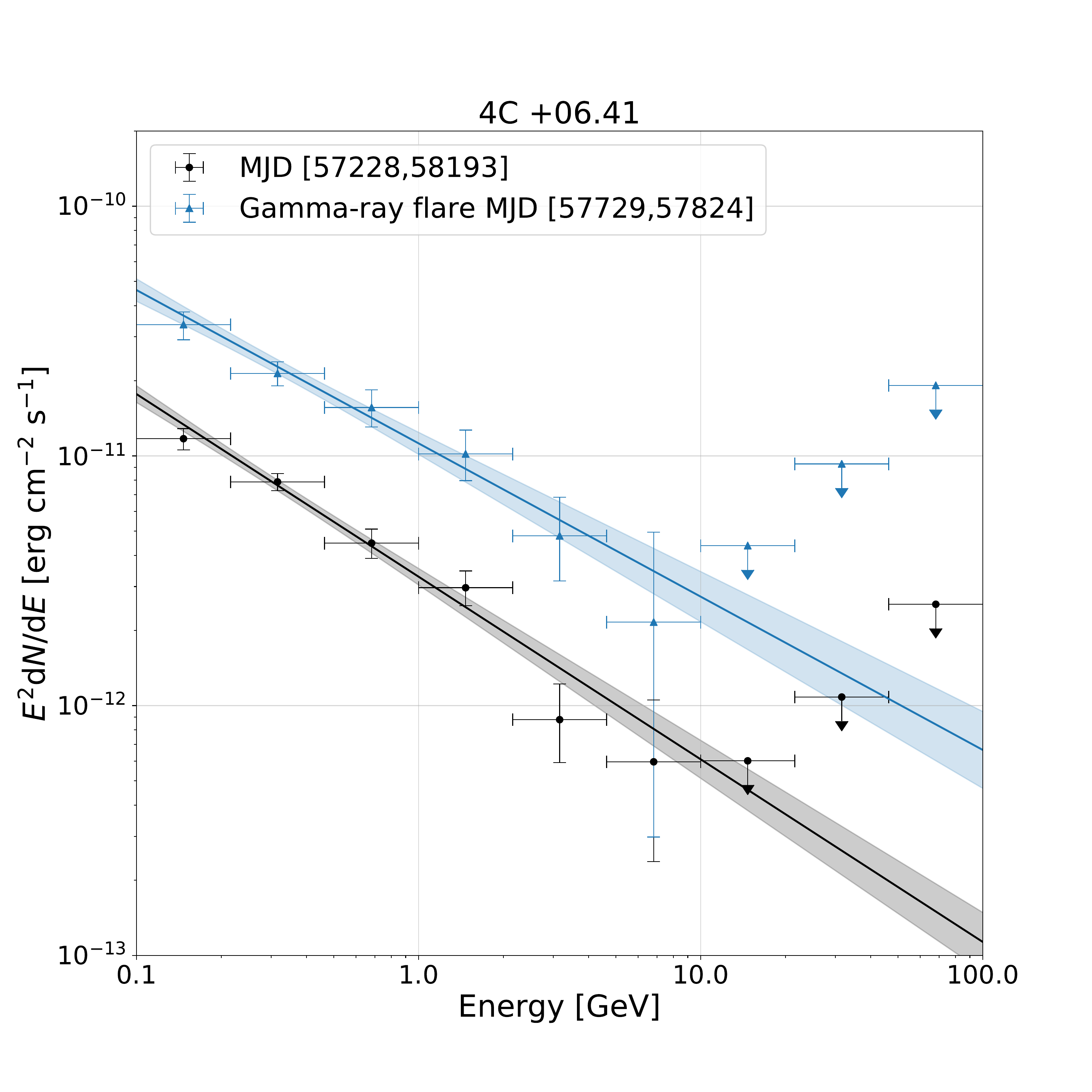}{0.5\textwidth}{(b)}
          }
  
  \caption{Spectral energy distributions of \GB and \CCCC. \Fermi-LAT data of the respective time ranges when each source is significantly detected are shown as black crosses, arrows indicate 95\% upper limits and the best-fit spectral model including statistical uncertainties is overlaid as a black band. 
Left: average spectrum of \GB compared to the spectrum during the 93-day gamma-ray excess coincident with the neutrino detection and the bright hard state during MJD 55753-56474. Right: average spectrum of \CCCC compared with the spectrum during the bright gamma-ray flare during MJD 57729-57824.}
  \label{fig:SED_J1040}
\end{figure*}

Fig.~\ref{fig:SED_J1040} (left) shows the SED averaged over the time window from MJD 54633 to 57227 (where we have a significant detection of \GB with $TS_{det}$ of 451) compared to the hard bright state and to the spectrum during the 93 days around the neutrino arrival time. The average gamma-ray emission is well modeled by a power law with $\gamma = 2.33\,\pm\,0.06$ and $N_{0} = (1.77\,\pm\,0.08)\,\times 10^{-12}$\,cm$^{-2}$\,s$^{-1}$\,MeV$^{-1}$. 
A likelihood ratio test similar to the one performed in Sec.~\ref{sec:TXS_LC} shows a hardening of the spectrum during the hard bright state at the $4.1\sigma$ level. Furthermore, we find that the spectral shape during the hard state favors a log-parabola instead of a power-law model, with best-fit spectral parameters $\alpha=2.03\pm 0.06$ and $\beta=0.10 \pm 0.03$ ($E_b$ is fixed to 1\,GeV) with a $19\sigma$ significance. We note that, during the bright hard state, there is an increase of at least a factor of 10 in the energy at which the high-energy component of the SED peaks.

During the 93-day window around the neutrino arrival time the source is brighter by a factor of 2.4 compared to the average integrated energy flux with a spectral shape compatible to the average one at one-sigma level. The best fit spectral parameters during this time are $\gamma = 2.43\,\pm\,0.11$ and $N_{0} = (3.76\,\pm\,0.55)\,\times 10^{-12}$\,cm$^{-2}$\,s$^{-1}$\,MeV$^{-1}$.

We do not find photons above 50\,GeV during the 9.6 years of Fermi-LAT observations, consistent with the source not being included in the 2FHL catalog. During the bright hard state we find 10 photons above 10\,GeV, which is compatible with $9.56$ expected photons from the average spectral shape with fitted normalization during the flare time. We do not find an excess of high-energy photons because the spectral change is mainly due to a lack in low-energy photons caused by a shift in the high-energy SED peak to higher energies. We find one photon with an energy larger than 10\,GeV during the flare at the neutrino arrival time, which is consistent with the expectation of $1.54$ photons obtained assuming the average spectral shape and the flux normalization during this flare.

\subsection{Gamma-ray light curve of \CCCC}

\begin{figure*}
  \centering
  \includegraphics[width=0.8\textwidth]{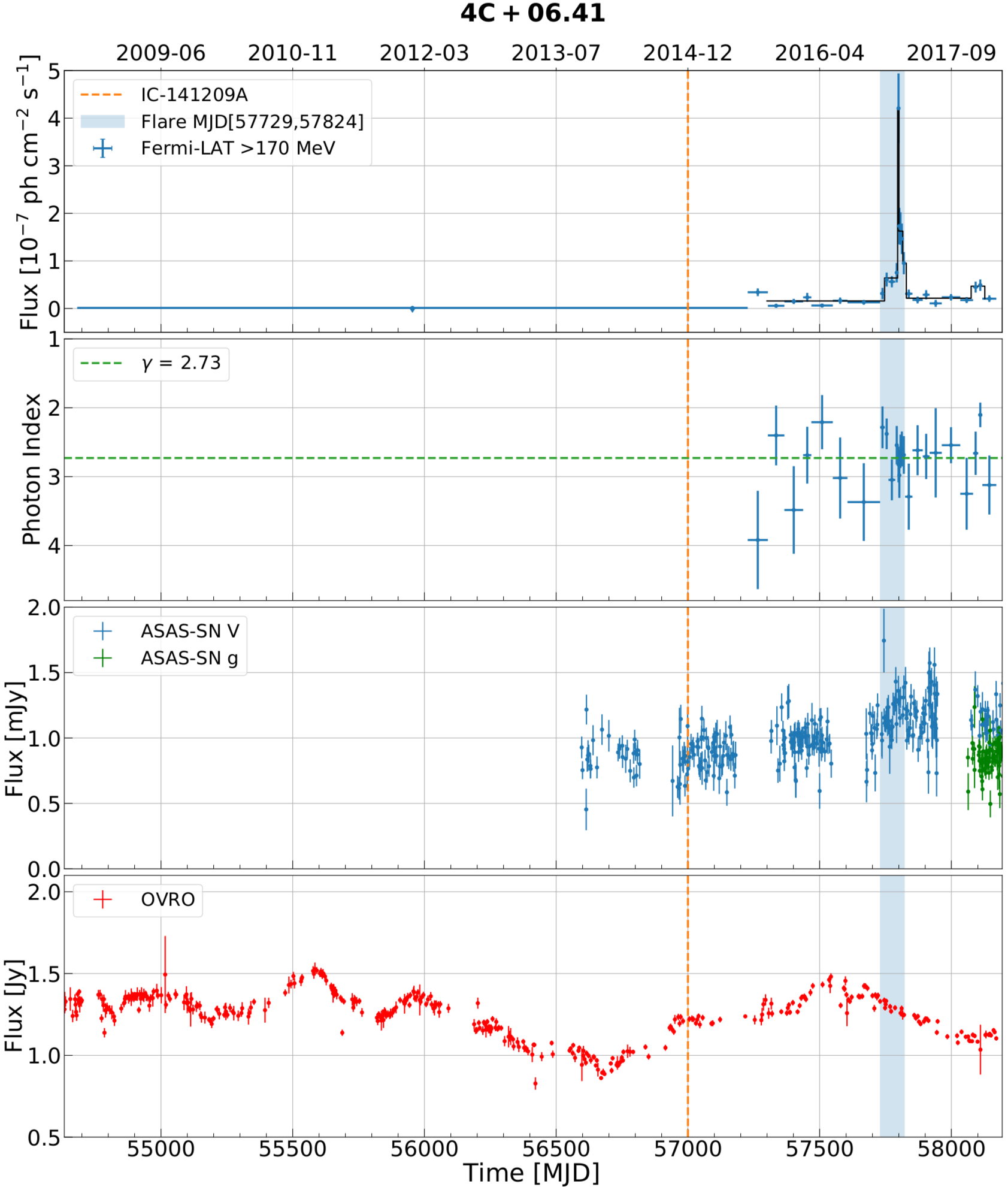}
  \caption{Adaptively-binned light curve for \CCCC. The first panel shows the gamma-ray flux above 170\,MeV including the Bayesian Block representation shown in black. The second panel shows the power-law spectral index. In the first time bin the source is not detected significantly, therefore a 95\% flux upper limit is shown in panel 1. In that case the spectral index cannot be fitted. The average spectral index is overlaid as a green line. The third panel shows optical data obtained from ASAS-SN, and the fourth panel shows radio data from OVRO. The arrival of \nuGB is indicated as an orange line.}
  \label{fig:LC_4C}
\end{figure*}

Fig.~\ref{fig:LC_4C} shows the adaptive binned light curve for \CCCC starting at the optimum energy of 170\,MeV beginning at MJD 57228. At earlier times no significant emission of the source is detected. The emission in the time window spanning from MJD 57228-58193 is well modeled by a power law with best-fit parameters of $\gamma = 2.73\,\pm\,0.05$, $N_{0} = (2.05\,\pm\,0.16)\,\times 10^{-13}$\,cm$^{-2}$\,s$^{-1}$\,MeV$^{-1}$ and reaches a $TS_{det}$ of 322.

At the arrival time of \nuGB the gamma-ray flux is below $1.44\,\times 10^{-9}$\,ph\,cm$^{-2}$\,s$^{-1}$ at 95\% confidence level, integrated between 300 MeV and 1 TeV. The source is in flaring state during a 95-day period between MJD 57729 to 57824 where it outshines \GB which is not significantly detected. During the flare the source follows a power-law spectrum with best-fit parameters $\gamma=2.61\,\pm\,0.07$ and $N_{0} = (7.01\,\pm\,0.75)\,\times 10^{-12}$\,cm$^{-2}$\,s$^{-1}$\,MeV$^{-1}$.

\subsection{Disentangling the Gamma-ray Emission}

The light curves presented in Fig.~\ref{fig:LC_J1040} and Fig.~\ref{fig:LC_4C} indicate that the gamma-ray emission from \GB dominates at earlier times, up to $\sim$mid-2015 when it entered a quiescent gamma-ray state, mostly below the detection sensitivity for the LAT. Mid-2015 is also the time around which \CCCC starts to emit a detectable gamma-ray flux. 
To prove that the temporal coincidence of the onset of the gamma-ray emission of \CCCC and the drop in gamma rays from \GB is not due to source confusion at low energies, we repeat the analysis at $>$1GeV (not shown). Here the improved LAT PSF minimizes the risk of source confusion. We retrieve similar results at high-energies, showing the robustness of our analysis.

\begin{figure}
  \centering
  
  \includegraphics[width=0.45\textwidth]{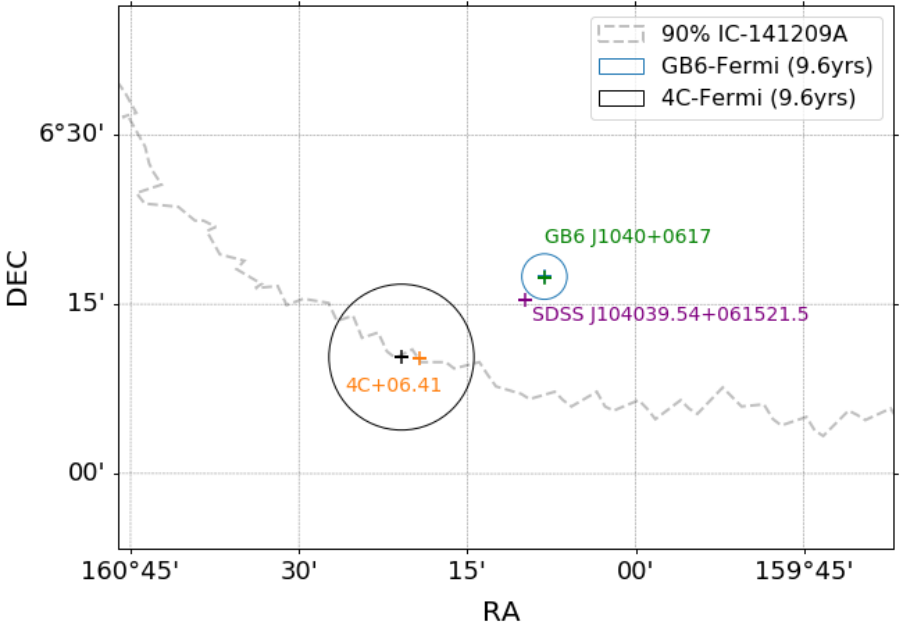}
  \caption{Gamma-ray best-fit positions: the blue and black circles indicate the 99\% containment radius of the gamma-ray positions of the two putative sources GB6-Fermi and 4C-Fermi. The 90\% neutrino uncertainty region is shown as dashed gray line for reference. Orange, green and violet crosses indicate the radio positions of the blazars located in the region.}
  \label{fig:Sky_Flare_9yrs}
\end{figure}

The radio-loud object \SDSS is located at RA, Dec = $160\fdg16475$, $6\fdg2558$, just 1 arcmin away from the radio position of \GB. Figure \ref{fig:Sky_Flare_9yrs} shows the best-fit gamma-ray localization, position and 99\% uncertainty, for two putative sources called GB6-Fermi and 4C-Fermi, using the best statistics available (full 9.6-years dataset). We find that the best-fit gamma-ray positions (blue cross for the first and black for the second) coincide well with the radio positions of \GB and \CCCC respectively (green and orange cross). \SDSS is located outside of both the 99\% uncertainty circles (blue circle) and is thus excluded as being responsible for the majority of the prolonged gamma-ray emission observed by the LAT. Adding another putative source at the radio position of \SDSS in our ROI does not significantly improve our model, yielding a significance of $TS_{det}=0$ for \SDSS. We calculate a 95\% flux upper limit for \SDSS of $8.8\times 10^{-10}$ph\,cm$^{-2}$\,s$^{-1}$ for a power-law spectral shape with index of 2.0 integrated over the energy range from 100\,MeV to 1\,TeV.

\begin{figure}
  \centering
  \includegraphics[width=0.45\textwidth]{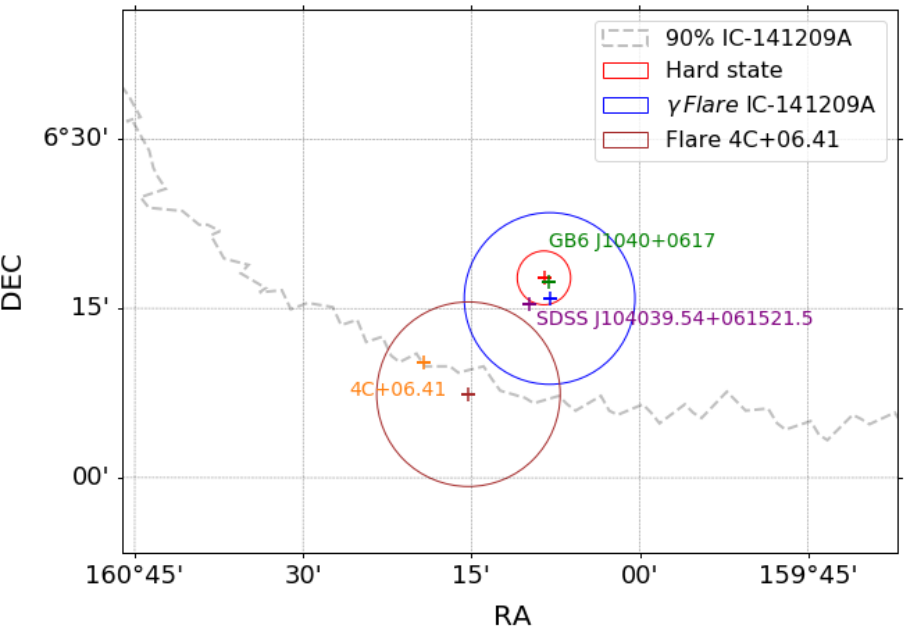}
  \caption{Gamma-ray flare positions: the bright hard state and modest flare at the arrival of \nuGB are shown in red and blue respectively, while the late flare attributed to \CCCC is shown in brown. The circles indicate the 99\% containment radius of the gamma-ray position. The 90\% neutrino uncertainty region is shown as dashed gray line.}
  \label{fig:Sky_Flare}
\end{figure}

As a sanity check, we ran a dedicated analysis for the flaring time intervals to derive the best-fit localization of the gamma-ray emission. We find that the bright hard state and the modest flare around the neutrino arrival time are consistent with the position of \GB while the most recent enhanced gamma-ray emission is positionally consistent with \CCCC (see Fig.~\ref{fig:Sky_Flare}). This is supported by the softer spectral shape observed during the most recent flare, matching well the one of \CCCC (see Fig.~\ref{fig:SED_J1040}). 
The association of the different flaring states to the two sources is supported by the temporal behavior of the two sources in the optical band (see Sec.~\ref{subsec:GB6MW}). 
While the lack of significant gamma-ray emission at the time of \nuGB does not exclude \CCCC from being the source of the neutrino, we focus here on \GB for closer multi-wavelength study, in light of the possible correlation between gamma-ray and neutrino emission.

\subsection{Multi-wavelength data collection}
\label{subsec:GB6MW}
Archival observations of the gamma-ray sources in the \nuGB region are available for several wavelengths.

Optical data in the V-band and g-band from the All-Sky Automated Survey for Supernovae \citep[ASAS-SN,][]{2014ApJ...788...48S,2017PASP..129j4502K} are processed by the fully automatic ASAS-SN pipeline using the ISIS image subtraction package \citep{alard98, alard00}. We remove science images by eye that are obviously affected by clouds. We then perform aperture photometry on the subtracted science image using the IRAF {\tt apphot} package, adding back in the flux from the reference image. The photometry is calibrated using the AAVSO Photometric All-Sky Survey (APASS; \citealp{henden15}). Additional V-band data from the Catalina Sky Survey ~\citep[CSS,][]{2009ApJ...696..870D} are available from the public database and are based on aperture photometry. R and g band light curves from the Palomar Transient Factory (PTF) are obtained from the IPAC archive~\citep{2014PASP..126..674L,2017PASP..129a4002M} and processed using forced PSF-fit photometry~\citep{2017PASP..129a4002M} on the subtracted images adding back the flux from the reference image. 
The long term optical light curve of \GB shown in the lower panel of Fig.~\ref{fig:LC_J1040} shows a similar flux variability pattern when compared to the gamma-ray light curve, including an excess coincident with the arrival of \nuGB.

The optical light curve of \CCCC, recorded by ASAS-SN and processed as outlined above, shows a mild excess around MJD 57800 in coincidence with the gamma-ray flare attributed to \CCCC (see Sec.~\ref{ssec:sec_lc_gb6}). The OVRO radio light curve shows a very slow rise starting around MJD 56700 and peaks almost one year before the gamma-ray flare.

\begin{figure*}
  \centering
  {\includegraphics[width=1.0\textwidth]{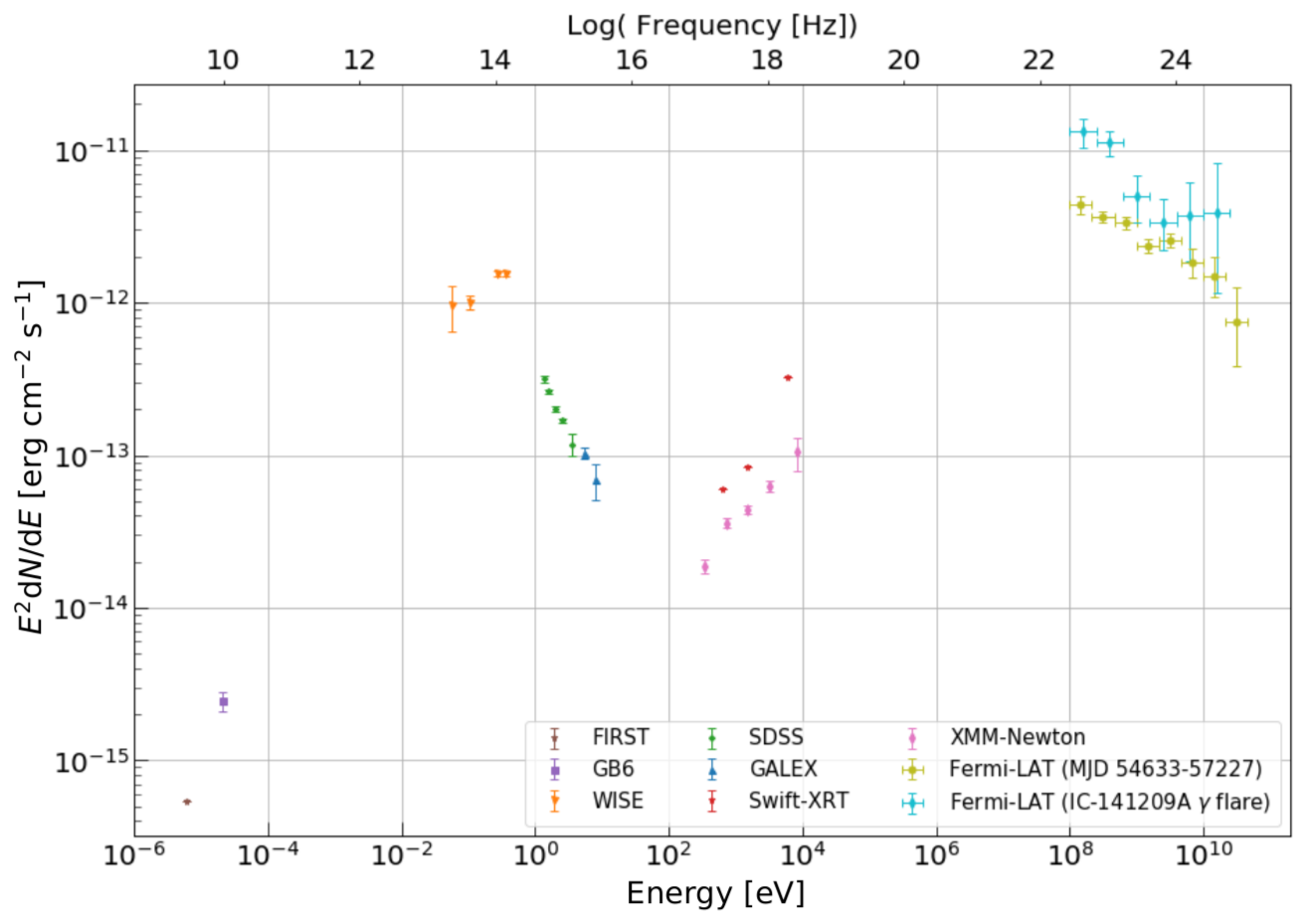}\label{fig:j1040_mwlsed}}
  
  \caption{Multi-wavelength SED in the observer's frame for \GB using archival data, which are not contemporaneous.}
  \label{fig:MWL_SED}
\end{figure*}

Fig.~\ref{fig:MWL_SED} shows an SED of \GB compiled from archival data. Note that these data are not contemporaneous. X-ray data are taken from the third \textit{XMM-Newton} serendipitous source catalog
\citep{2016A&A...590A...1R} and the \textit{Swift} XRT point source catalog~\citep{2014ApJS..210....8E}. We observe a flux difference in the \textit{Swift}-XRT and XMM-\textit{Newton} data, which we attribute to different observation periods. XMM-\textit{Newton} data were collected in 2003 May while \textit{Swift}-XRT observed the source between 2007 and 2011. Radio data come from the GB6 catalog of radio sources~\citep{1996ApJS..103..427G} and the FIRST survey~\citep{2015ApJ...801...26H}. Optical data are obtained from SDSS~\citep{GB_redshift} and far and near UV observations from \emph{GALEX}~\citep{2017ApJS..230...24B}. Infrared data are obtained from \emph{WISE}~\citep{2010AJ....140.1868W}. The SED shows the typical two-hump structure with the high-energy peak at $\sim 100$\,MeV and the low-energy peak in the infrared around $0.1$\,eV, which makes it a low-peaked synchrotron source.

\section{Conclusions}

High-energy neutrino production in blazars is accompanied by the production of gamma rays at similar energies. While the neutrinos escape from the system, the gamma rays can interact and cascade down to lower energies. Sources bright at GeV gamma rays are capable of accelerating particles to high energies and thus may be good tracers for neutrino emission. This paper presents a detailed characterization of the gamma-ray behavior for the potential electromagnetic counterparts spatially consistent with two well-reconstructed IceCube neutrinos put into a multi-frequency perspective.
\subsection{\nuTXS gamma-ray counterpart}
In \citet{MWScience} \TXS was suggested as the counterpart of \nuTXS. The refined gamma-ray analysis presented here confirms that at the time of the neutrino detection this blazar was undergoing a major, prolonged gamma-ray outburst phase without significant spectral variations. However, during the time of the 2014/15 neutrino flare reported in \citet{ICScience}, we find neither an excess of gamma rays nor a significant gamma-ray spectral change with respect to the average. This could point to absorption of the gamma rays~\citep{2018arXiv180705113L} or to an increase in the injection of protons, potentially explainable by hybrid models as the one proposed by e.g.~\citealt{Rodrigues:2018tku,Murase:2018iyl}. \citet{Reimer:2018vvw} have conducted a detailed investigation of the electromagnetic signal expected for photo-hadronically produced neutrinos in 2014/15 by \TXS. Comparing simulations to the observed data, they show that the link between gamma rays and neutrinos in this blazar may not be trivial. They derive the conclusion that in most of the considered scenarios the observed high-energy photons and neutrinos may be not casually connected. 

The bright gamma-ray flare in 2017 coincident with \nuTXS shows significant variations on short time scales (see Fig.~\ref{fig:LC_TXS_zoom}). While the source is detected with high confidence on daily time scales, significant variations are only found on a weekly timescale (by the Bayesian Block algorithm). However, fast variability on 1-day time scale was found in $>100$\,GeV gamma rays by MAGIC~\citep{Ahnen:2018mvi} and points to a compact emission region.

\subsection{\nuGB gamma-ray counterpart candidates}

If \nuGB is astrophysical in origin, then the low-synchrotron peaked gamma-ray blazar \GB appears to be the most likely counterpart, assuming a direct correlation between the gamma-ray and neutrino emission. Under that assumption the neighboring FSRQs \CCCC and \SDSS are less favored as the likely neutrino counterpart, because no significant high-energy emission was detected at the arrival of \nuGB. However, in models that assume a different scaling of the neutrino flux with the electromagnetic emission, \CCCC and \SDSS may be considered as potential neutrino counterparts \citep[e.g.][]{1991PhRvL..66.2697S,2016PhRvL.116g1101M,Reimer:2018vvw}. 

The variability pattern of \GB displays major and minor flaring episodes in both gamma-ray and optical wavelengths. At the detection time of \nuGB the blazar showed an increase in the gamma-ray flux over 93 days, with respect to the 9.6 years averaged flux (Fig.~\ref{fig:LC_J1040}, panel 1). A Bayesian block analysis confirms the flaring activity. However, the bulk of the $>300$\,MeV gamma-ray energy output is observed during the 721-day long high-flux state before the neutrino detection, as evidenced by the light curve shown in Fig.~\ref{fig:LC_J1040} (panel 1). The source entered a lower active state roughly 100 days after the neutrino arrival.

\begin{figure}
  \centering
  \includegraphics[width=0.5\textwidth]{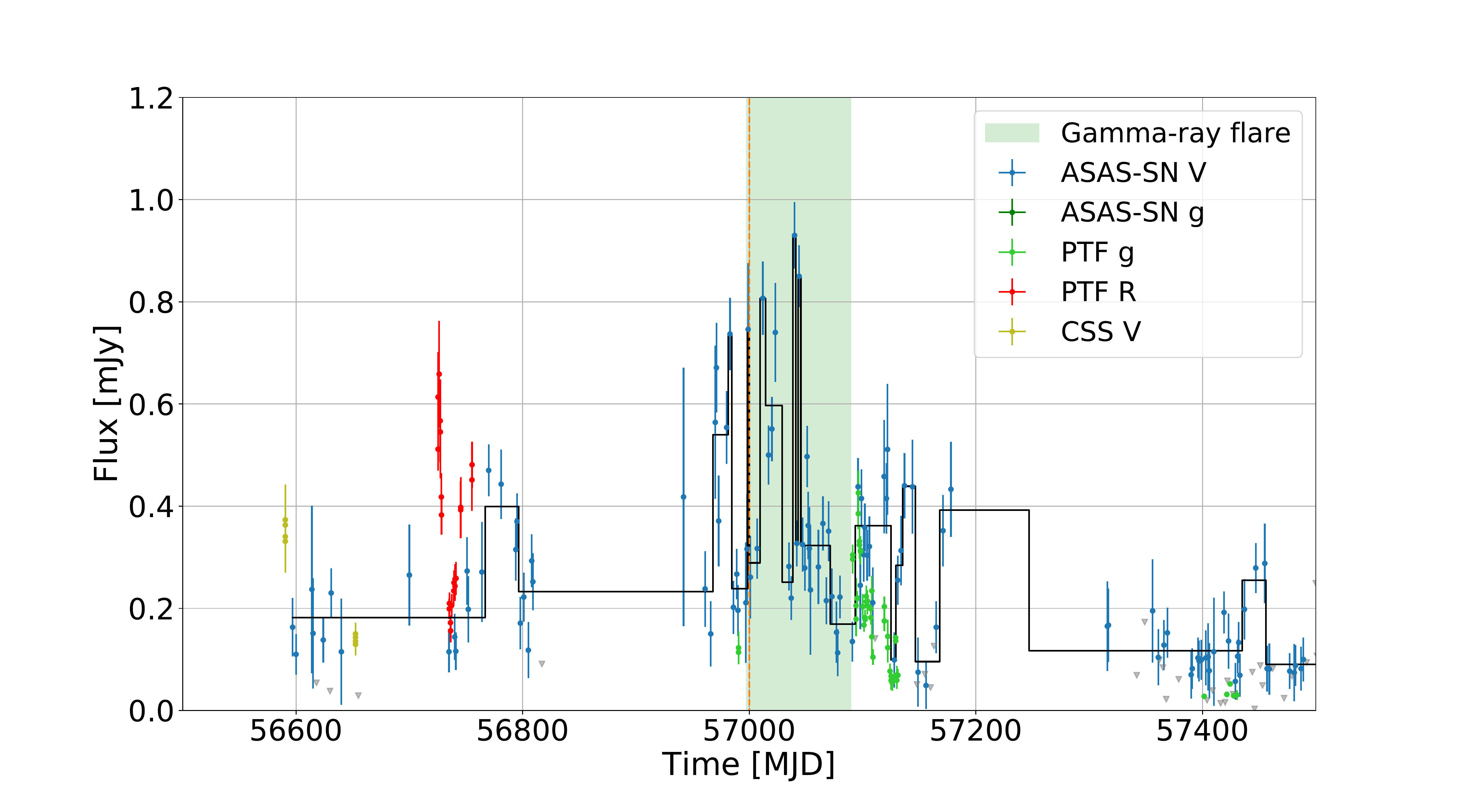}
  \caption{Zoomed in optical light curve of \GB around the arrival time of \nuGB (shown in orange). The green shaded region indicates the gamma-ray flare in coincidence with the neutrino arrival time (see Fig.~\ref{fig:LC_J1040}). In black is shown the Bayesian Block representation for the ASAS-SN V-band data set.}
  \label{fig:LC_J1040_zoom}
\end{figure}

Enhanced activity contemporaneous to \nuGB is also supported by an overall steady increase in the object's flux in the optical band (Fig.~\ref{fig:LC_J1040}, panel 3). Simultaneous ASAS-SN observations confirm that, at the neutrino detection time, the blazar's optical flux was higher than average, and displayed the second brightest historical value (while the record-holder was the optical flare coincident with the bright, hard gamma-ray state). A zoom-in of the optical light curve around the neutrino arrival time (Fig.~\ref{fig:LC_J1040_zoom}) shows an increase compared to the low state by almost a factor of 10. We do not find a hint for fast gamma-ray variability during this period, while short-timescale flux variations were evident in the optical band. This could be compatible with the proton-synchrotron scenario discussed in~\citet{2018arXiv180711069Z}. However, low gamma-ray statistics and the lack of polarization information prevents us from probing the model predictions.

Assuming a redshift of 0.73, with the caveats mentioned earlier, the rest-frame energetics of \GB are similar to those of \TXS.  The average gamma-ray luminosity between 100\,MeV and 100\,GeV was $4.1\times10^{46}$\,erg/s, which is 1.5 times larger than \TXS~\citep{MWScience}\footnote{Note that with an average redshift of \Fermi-LAT BL Lacs of $z=0.3$ the estimated luminosity of \GB reduces to $4.9\times10^{45}$\,erg/s.}. The modest flare at the arrival of \nuGB had a luminosity of $8.9\times10^{46}$\,erg/s.  

We perform a rough estimate of the expected neutrino event rate assuming that the average neutrino flux reaches at most the level of the gamma-ray flux. In hybrid models the X-ray flux should be dominated by the hadronic component~\citep[e.g. ][]{Gao:2018mnu,Keivani:2018rnh} and therefore for our rough estimate we assume that the neutrino flux has to be higher than the X-ray flux. We assume a peak gamma-ray flux of $E^2$ d$N$/d$E \approx 5 \times 10^{-12}$\,erg\,cm$^{-2}$\,s$^{-1}$, a minimum X-ray flux of $E^2$ d$N$/d$E \approx 2 \times 10^{-14}$\,erg\,cm$^{-2}$\,s$^{-1}$ and a neutrino spectral shape of $E^{-2}$. We use the IceCube effective area at the declination of \GB (which is similar to the effective area published in \citet{ICScience} for \TXS using the 86-string configuration of 2015 May 18 to 2017 October 31). We find an optimistic expected number of neutrinos between 100\,GeV and 10\,PeV of 1.8 events per year and 0.7 events above 100\,TeV for a neutrino flux as high as the peak gamma-ray flux and a pessimistic number of 0.01 per year between 100\,GeV and 10\,PeV and 0.003 per year above 100\,TeV. We note that those numbers are only rough estimates and are highly model dependent. An expectation value much smaller than one is compatible with the detection of a single high-energy event due to the Eddington bias discussed in~\citet{Strotjohann:2018ufz}. This shows that \GB is 
a plausible neutrino-source candidate and motivates the search for TeV neutrinos at this source position similar to the one performed at the position of \TXS~\citep{ICScience}, which is in preparation.

\subsection{Gamma-ray blazars as neutrino source population}
Only two of the 37 well reconstructed high-energy neutrino events satisfying the realtime trigger criteria are found to be positionally consistent with sources in the \Fermi-LAT energy range. 
Finding two out of $18.5\pm1.8$ events\footnote{Here we assume a signal fraction of our sample of $45-55$\%. The spread includes systematic uncertainties due to uncertainties in the assumed neutrino spectral shape and uncertainties in the signal fraction introduced by removing badly reconstructed events from the initial stream.} originating from \Fermi blazars is consistent with the blazar stacking limit performed in~\citet{Aartsen:2016lir} constraining the blazar contribution to the measured diffuse neutrino flux to $<30\%$.

Both \TXS and \GB share similar properties: they belong to the BL Lac class, the former to the sub-sample of ISP and the latter to the one of LSP sources, and have comparable gamma-ray luminosities. Moreover, they are located at a similar declination, near the equatorial plane, which is viewed along the horizon from the South Pole. This is the sky region for which IceCube is most sensitive to high-energy neutrinos. While we do not have significant evidence that \nuGB is associated with any of the gamma-ray objects identified in its vicinity, our multi-wavelength study suggests that based on its gamma-ray properties, \GB 
remains a plausible candidate for being a gamma-ray counterpart to the neutrino event.
However, given the currently limited knowledge of the blazar jet properties and acceleration mechanisms leading to an uncertainty in the scaling of the neutrino flux with the electromagnetic emission and the lack of simultaneous multi-wavelength data, \SDSS and \CCCC cannot be ruled out as possible counterparts of \nuGB.

This work points to the importance of broad-band multi-wavelength and multi-messenger data to provide us with a more complete understanding of candidate neutrino counterparts. While gamma rays are closest in energy to the neutrinos of interest, lower energy photons produced in cascades also have to be considered as tracers of increased hadronic activity of the source. These will be a crucial ingredient in future searches for neutrino emitters, and hence cosmic-ray source populations.

\begin{acknowledgements}

The authors thank Matthias Kadler, Anatoli Fedynitch and Shan Gao for fruitful discussions. SG and AF are supported by the
Initiative and Networking Fund of the Helmholtz Association. SB has received support by the NASA NPP Fellowship. 

\textbf{Fermi-LAT:} The \textit{Fermi} LAT Collaboration acknowledges generous ongoing support
from a number of agencies and institutes that have supported both the
development and the operation of the LAT as well as scientific data analysis.
These include the National Aeronautics and Space Administration and the
Department of Energy in the United States, the Commissariat \`a l'Energie Atomique
and the Centre National de la Recherche Scientifique / Institut National de Physique
Nucl\'eaire et de Physique des Particules in France, the Agenzia Spaziale Italiana
and the Istituto Nazionale di Fisica Nucleare in Italy, the Ministry of Education,
Culture, Sports, Science and Technology (MEXT), High Energy Accelerator Research
Organization (KEK) and Japan Aerospace Exploration Agency (JAXA) in Japan, and
the K.~A.~Wallenberg Foundation, the Swedish Research Council and the
Swedish National Space Board in Sweden.
 
Additional support for science analysis during the operations phase is gratefully
acknowledged from the Istituto Nazionale di Astrofisica in Italy and the Centre
National d'\'Etudes Spatiales in France. This work performed in part under DOE
Contract DE-AC02-76SF00515.


\textbf{ASAS-SN:} The ASAS-SN team thanks the Las Cumbres Observatory and its staff for its continuing support of the ASAS-SN project.  ASAS-SN is supported by the Gordon and Betty Moore Foundation through grant GBMF5490 to the Ohio State University and NSF grant AST-1515927. Development of ASAS-SN has been supported by NSF grant AST-0908816, the Mt. Cuba Astronomical Foundation, the Center for Cosmology and AstroParticle Physics at the Ohio State University, the Chinese Academy of Sciences South America Center for Astronomy (CASSACA), the Villum Foundation, and George Skestos.
J.F.B. is supported by NSF Grant No. PHY-1714479.

\textbf{IceCube:} The IceCube collaboration provided the analysis of \nuGB via significant contributions by Claudio Kopper.  
The IceCube collaboration gratefully acknowledge the support from
the following agencies and institutions: USA -- U.S. National Science Foundation-Office of Polar Programs,
U.S. National Science Foundation-Physics Division,
Wisconsin Alumni Research Foundation,
Center for High Throughput Computing (CHTC) at the University of Wisconsin-Madison,
Open Science Grid (OSG),
Extreme Science and Engineering Discovery Environment (XSEDE),
U.S. Department of Energy-National Energy Research Scientific Computing Center,
Particle astrophysics research computing center at the University of Maryland,
Institute for Cyber-Enabled Research at Michigan State University,
and Astroparticle physics computational facility at Marquette University;
Belgium -- Funds for Scientific Research (FRS-FNRS and FWO),
FWO Odysseus and Big Science programmes,
and Belgian Federal Science Policy Office (Belspo);
Germany -- Bundesministerium f\"ur Bildung und Forschung (BMBF),
Deutsche Forschungsgemeinschaft (DFG),
Helmholtz Alliance for Astroparticle Physics (HAP),
Initiative and Networking Fund of the Helmholtz Association,
Deutsches Elektronen Synchrotron (DESY),
and High Performance Computing cluster of the RWTH Aachen;
Sweden -- Swedish Research Council,
Swedish Polar Research Secretariat,
Swedish National Infrastructure for Computing (SNIC),
and Knut and Alice Wallenberg Foundation;
Australia -- Australian Research Council;
Canada -- Natural Sciences and Engineering Research Council of Canada,
Calcul Qu\'ebec, Compute Ontario, Canada Foundation for Innovation, WestGrid, and Compute Canada;
Denmark -- Villum Fonden, Danish National Research Foundation (DNRF);
New Zealand -- Marsden Fund;
Japan -- Japan Society for Promotion of Science (JSPS)
and Institute for Global Prominent Research (IGPR) of Chiba University;
Korea -- National Research Foundation of Korea (NRF); Switzerland -- Swiss National Science Foundation (SNSF).

\textbf{Others:} The CSS survey is funded by the National Aeronautics and Space
Administration under Grant No. NNG05GF22G issued through the Science
Mission Directorate Near-Earth Objects Observations Program.  The CRTS
survey is supported by the U.S.~National Science Foundation under
grants AST-0909182. This research has made use of data from the OVRO 40-m monitoring program (Richards, J. L. et al. 2011, ApJS, 194, 29) which is supported in part by NASA grants NNX08AW31G, NNX11A043G, and NNX14AQ89G and NSF grants AST-0808050 and AST-1109911.

\end{acknowledgements}
\software{Fermi-LAT ScienceTools (v11r5p3), SIS (Alard $\&$Lupton 1998; Alard 2000), IRAF (Tody 1986, Tody 1993), Fermipy \citep{Wood:2017yyb},
Astropy \citep[]{astropy:2013,astropy:2018}}

\bibliography{literature}

\begin{thebibliography}{}
\expandafter\ifx\csname natexlab\endcsname\relax\def\natexlab#1{#1}\fi
\providecommand{\url}[1]{\href{#1}{#1}}

\bibitem[{Aartsen {et~al.}(2013)Aartsen, Abbasi, Abdouand,
  {et~al.}}]{Aartsen:2013jdh}
Aartsen, M.~G., Abbasi, R., Abdouand, Y., {et~al.} 2013, Science, 342, 1242856

\bibitem[{Aartsen {et~al.}(2018{\natexlab{a}})Aartsen, Ackermann, Adams,
  {et~al.}}]{MWScience}
Aartsen, M.~G., Ackermann, M., Adams, J., {et~al.} 2018{\natexlab{a}}, Science,
  361, eaat1378

\bibitem[{Aartsen {et~al.}(2018{\natexlab{b}})Aartsen, Ackermann, Adams,
  {et~al.}}]{ICScience}
---. 2018{\natexlab{b}}, Science, 361, 147

\bibitem[{Aartsen {et~al.}(2015)Aartsen, Ackermann, Adams, Aguilar, Ahlers,
  Ahrens, Altmann, Anderson, Arguelles, Arlen, Auffenberg, Bai, Barwick, Baum,
  Bay, Beatty, Becker~Tjus, Becker, BenZvi, Berghaus, Berley, Bernardini,
  Bernhard, Besson, Binder, Bindig, Bissok, Blaufuss, Blumenthal, Boersma,
  Bohm, Bos, Bose, B\"oser, Botner, Brayeur, Bretz, Brown, Buzinsky, Casey,
  Casier, Cheung, Chirkin, Christov, Christy, Clark, Classen, Clevermann,
  Coenders, Cowen, Cruz~Silva, Danninger, Daughhetee, Davis, Day, de~Andr\'e,
  De~Clercq, De~Ridder, Desiati, de~Vries, de~With, DeYoung, D\'{\i}az-V\'elez,
  Dunkman, Eagan, Eberhardt, Eichmann, Eisch, Euler, Evenson, Fadiran, Fazely,
  Fedynitch, Feintzeig, Felde, Feusels, Filimonov, Finley, Fischer-Wasels,
  Flis, Franckowiak, Frantzen, Fuchs, Gaisser, Gaior, Gallagher, Gerhardt,
  Gier, Gladstone, Gl\"usenkamp, Goldschmidt, Golup, Gonzalez, Goodman, G\'ora,
  Grant, Gretskov, Groh, Gro\ss{}, Ha, Haack, Haj~Ismail, Hallen, Hallgren,
  Halzen, Hanson, Hebecker, Heereman, Heinen, Helbing, Hellauer, Hellwig,
  Hickford, Hill, Hoffman, Hoffmann, Homeier, Hoshina, Huang, Huelsnitz, Hulth,
  Hultqvist, Hussain, Ishihara, Jacobi, Jacobsen, Jagielski, Japaridze, Jero,
  Jlelati, Jurkovic, Kaminsky, Kappes, Karg, Karle, Kauer, Keivani, Kelley,
  Kheirandish, Kiryluk, Kl\"as, Klein, K\"ohne, Kohnen, Kolanoski, Koob,
  K\"opke, Kopper, Kopper, Koskinen, Kowalski, Kriesten, Krings, Kroll, Kroll,
  Kunnen, Kurahashi, Kuwabara, Labare, Larsen, Larson, Lesiak-Bzdak, Leuermann,
  Leute, L\"unemann, Madsen, Maggi, Maruyama, Mase, Matis, Maunu, McNally,
  Meagher, Medici, Meli, Meures, Miarecki, Middell, Middlemas, Milke, Miller,
  Mohrmann, Montaruli, Morse, Nahnhauer, Naumann, Niederhausen, Nowicki,
  Nygren, Obertacke, Odrowski, Olivas, Omairat, O'Murchadha, Palczewski, Paul,
  Penek, Pepper, P\'erez de~los Heros, Pfendner, Pieloth, Pinat, Posselt,
  Price, Przybylski, P\"utz, Quinnan, R\"adel, Rameez, Rawlins, Redl, Rees,
  Reimann, Relich, Resconi, Rhode, Richman, Riedel, Robertson, Rodrigues,
  Rongen, Rott, Ruhe, Ruzybayev, Ryckbosch, Saba, Sander, Sandroos, Santander,
  Sarkar, Schatto, Scheriau, Schmidt, Schmitz, Schoenen, Sch\"oneberg,
  Sch\"onwald, Schukraft, Schulte, Schulz, Seckel, Sestayo, Seunarine,
  Shanidze, Smith, Soldin, Spiczak, Spiering, Stamatikos, Stanev, Stanisha,
  Stasik, Stezelberger, Stokstad, St\"o\ss{}l, Strahler, Str\"om, Strotjohann,
  Sullivan, Taavola, Taboada, Tamburro, Tepe, Ter-Antonyan, Terliuk, Te\ifmmode
  \check{s}\else \v{s}\fi{}i\ifmmode~\acute{c}\else \'{c}\fi{}, Tilav, Toale,
  Tobin, Tosi, Tselengidou, Unger, Usner, Vallecorsa, van Eijndhoven,
  Vandenbroucke, van Santen, Vehring, Voge, Vraeghe, Walck, Wallraff, Weaver,
  Wellons, Wendt, Westerhoff, Whelan, Whitehorn, Wichary, Wiebe, Wiebusch,
  Williams, Wissing, Wolf, Wood, Woschnagg, Xu, Xu, Yanez, Yodh, Yoshida,
  Zarzhitsky, Ziemann, Zierke, \& Zoll}]{PhysRevD.91.022001}
---. 2015, Phys. Rev. D, 91, 022001.
\newblock \url{https://link.aps.org/doi/10.1103/PhysRevD.91.022001}

\bibitem[{{Aartsen} {et~al.}(2015){Aartsen}, {Ackermann}, {Adams}, {Aguilar},
  {Ahlers}, {Ahrens}, {Altmann}, {Anderson}, {Archinger}, {Arguelles}, \&
  et~al.}]{2015ApJ...807...46A}
{Aartsen}, M.~G., {Ackermann}, M., {Adams}, J., {et~al.} 2015, \apj, 807, 46

\bibitem[{Aartsen {et~al.}(2016)}]{Aartsen:2016xlq}
Aartsen, M.~G., {et~al.} 2016, Astrophys. J., 833, 3

\bibitem[{Aartsen {et~al.}(2017{\natexlab{a}})}]{Aartsen:2016oji}
---. 2017{\natexlab{a}}, Astrophys. J., 835, 151

\bibitem[{Aartsen {et~al.}(2017{\natexlab{b}})}]{2017APh....92...30A}
---. 2017{\natexlab{b}}, Astroparticle Physics, 92, 30

\bibitem[{Aartsen {et~al.}(2017{\natexlab{c}})}]{2017JInst..12P3012A}
---. 2017{\natexlab{c}}, Journal of Instrumentation, 12, P03012

\bibitem[{Aartsen {et~al.}(2017{\natexlab{d}})}]{Aartsen:2016lir}
---. 2017{\natexlab{d}}, Astrophys. J., 835, 45

\bibitem[{{Abeysekara} {et~al.}(2018){Abeysekara}, {Archer}, {Benbow}, {Bird},
  {Brill}, {Brose}, {Buckley}, {Christiansen}, {Chromey}, {Daniel}, {Falcone},
  {The VERITAS Collaboration}, \& {Kaur}}]{2018ApJ...861L..20A}
{Abeysekara}, A.~U., {Archer}, A., {Benbow}, W., {et~al.} 2018, \apjl, 861, L20

\bibitem[{{Abolfathi} {et~al.}(2018){Abolfathi}, {Aguado}, {Aguilar}, {Allende
  Prieto}, {Almeida}, {Ananna}, {Anders}, {Anderson}, {Andrews}, {Anguiano}, \&
  et~al.}]{GB_redshift}
{Abolfathi}, B., {Aguado}, D.~S., {Aguilar}, G., {et~al.} 2018, \apjs, 235, 42

\bibitem[{{Acero} {et~al.}(2015){Acero}, {Ackermann}, {Ajello}, {Albert},
  {Atwood}, {Axelsson}, {Baldini}, {Ballet}, {Barbiellini}, {Bastieri},
  {Belfiore}, {Bellazzini}, {Bissaldi}, {Blandford}, {Bloom}, {Bogart},
  {Bonino}, {Bottacini}, {Bregeon}, {Britto}, {Bruel}, {Buehler}, {Burnett},
  {Buson}, {Caliandro}, {Cameron}, {Caputo}, {Caragiulo}, {Caraveo},
  {Casandjian}, {Cavazzuti}, {Charles}, {Chaves}, {Chekhtman}, {Cheung},
  {Chiang}, {Chiaro}, {Ciprini}, {Claus}, {Cohen-Tanugi}, {Cominsky}, {Conrad},
  {Cutini}, {D'Ammando}, {de Angelis}, {DeKlotz}, {de Palma}, {Desiante},
  {Digel}, {Di Venere}, {Drell}, {Dubois}, {Dumora}, {Favuzzi}, {Fegan},
  {Ferrara}, {Finke}, {Franckowiak}, {Fukazawa}, {Funk}, {Fusco}, {Gargano},
  {Gasparrini}, {Giebels}, {Giglietto}, {Giommi}, {Giordano}, {Giroletti},
  {Glanzman}, {Godfrey}, {Grenier}, {Grondin}, {Grove}, {Guillemot}, {Guiriec},
  {Hadasch}, {Harding}, {Hays}, {Hewitt}, {Hill}, {Horan}, {Iafrate}, {Jogler},
  {J{\'o}hannesson}, {Johnson}, {Johnson}, {Johnson}, {Johnson}, {Kamae},
  {Kataoka}, {Katsuta}, {Kuss}, {La Mura}, {Landriu}, {Larsson}, {Latronico},
  {Lemoine-Goumard}, {Li}, {Li}, {Longo}, {Loparco}, {Lott}, {Lovellette},
  {Lubrano}, {Madejski}, {Massaro}, {Mayer}, {Mazziotta}, {McEnery},
  {Michelson}, {Mirabal}, {Mizuno}, {Moiseev}, {Mongelli}, {Monzani},
  {Morselli}, {Moskalenko}, {Murgia}, {Nuss}, {Ohno}, {Ohsugi}, {Omodei},
  {Orienti}, {Orlando}, {Ormes}, {Paneque}, {Panetta}, {Perkins},
  {Pesce-Rollins}, {Piron}, {Pivato}, {Porter}, {Racusin}, {Rando}, {Razzano},
  {Razzaque}, {Reimer}, {Reimer}, {Reposeur}, {Rochester}, {Romani},
  {Salvetti}, {S{\'a}nchez-Conde}, {Saz Parkinson}, {Schulz}, {Siskind},
  {Smith}, {Spada}, {Spandre}, {Spinelli}, {Stephens}, {Strong}, {Suson},
  {Takahashi}, {Takahashi}, {Tanaka}, {Thayer}, {Thayer}, {Thompson},
  {Tibaldo}, {Tibolla}, {Torres}, {Torresi}, {Tosti}, {Troja}, {Van Klaveren},
  {Vianello}, {Winer}, {Wood}, {Wood}, {Zimmer}, \& {Fermi-LAT
  Collaboration}}]{2015ApJS..218...23A}
{Acero}, F., {Ackermann}, M., {Ajello}, M., {et~al.} 2015, \apjs, 218, 23

\bibitem[{{Ackermann} {et~al.}(2015){Ackermann}, {Ajello}, {Atwood}, {Baldini},
  {Ballet}, {Barbiellini}, {Bastieri}, {Becerra Gonzalez}, {Bellazzini},
  {Bissaldi}, {Blandford}, {Bloom}, {Bonino}, {Bottacini}, {Brandt}, {Bregeon},
  {Britto}, {Bruel}, {Buehler}, {Buson}, {Caliandro}, {Cameron}, {Caragiulo},
  {Caraveo}, {Carpenter}, {Casandjian}, {Cavazzuti}, {Cecchi}, {Charles},
  {Chekhtman}, {Cheung}, {Chiang}, {Chiaro}, {Ciprini}, {Claus},
  {Cohen-Tanugi}, {Cominsky}, {Conrad}, {Cutini}, {D'Abrusco}, {D'Ammando}, {de
  Angelis}, {Desiante}, {Digel}, {Di Venere}, {Drell}, {Favuzzi}, {Fegan},
  {Ferrara}, {Finke}, {Focke}, {Franckowiak}, {Fuhrmann}, {Fukazawa},
  {Furniss}, {Fusco}, {Gargano}, {Gasparrini}, {Giglietto}, {Giommi},
  {Giordano}, {Giroletti}, {Glanzman}, {Godfrey}, {Grenier}, {Grove},
  {Guiriec}, {Hewitt}, {Hill}, {Horan}, {Itoh}, {J{\'o}hannesson}, {Johnson},
  {Johnson}, {Kataoka}, {Kawano}, {Krauss}, {Kuss}, {La Mura}, {Larsson},
  {Latronico}, {Leto}, {Li}, {Li}, {Longo}, {Loparco}, {Lott}, {Lovellette},
  {Lubrano}, {Madejski}, {Mayer}, {Mazziotta}, {McEnery}, {Michelson},
  {Mizuno}, {Moiseev}, {Monzani}, {Morselli}, {Moskalenko}, {Murgia}, {Nuss},
  {Ohno}, {Ohsugi}, {Ojha}, {Omodei}, {Orienti}, {Orlando}, {Paggi}, {Paneque},
  {Perkins}, {Pesce-Rollins}, {Piron}, {Pivato}, {Porter}, {Rain{\`o}},
  {Rando}, {Razzano}, {Razzaque}, {Reimer}, {Reimer}, {Romani}, {Salvetti},
  {Schaal}, {Schinzel}, {Schulz}, {Sgr{\`o}}, {Siskind}, {Sokolovsky}, {Spada},
  {Spandre}, {Spinelli}, {Stawarz}, {Suson}, {Takahashi}, {Takahashi},
  {Tanaka}, {Thayer}, {Thayer}, {Tibaldo}, {Torres}, {Torresi}, {Tosti},
  {Troja}, {Uchiyama}, {Vianello}, {Winer}, {Wood}, \&
  {Zimmer}}]{2015ApJ...810...14A}
{Ackermann}, M., {Ajello}, M., {Atwood}, W.~B., {et~al.} 2015, \apj, 810, 14

\bibitem[{{Ackermann} {et~al.}(2016){Ackermann}, {Ajello}, {Atwood}, {Baldini},
  {Ballet}, {Barbiellini}, {Bastieri}, {Becerra Gonzalez}, {Bellazzini},
  {Bissaldi}, {Blandford}, {Bloom}, {Bonino}, {Bottacini}, {Brandt}, {Bregeon},
  {Bruel}, {Buehler}, {Buson}, {Caliandro}, {Cameron}, {Caputo}, {Caragiulo},
  {Caraveo}, {Cavazzuti}, {Cecchi}, {Charles}, {Chekhtman}, {Cheung}, {Chiang},
  {Chiaro}, {Ciprini}, {Cohen}, {Cohen-Tanugi}, {Cominsky}, {Conrad}, {Cuoco},
  {Cutini}, {D'Ammando}, {de Angelis}, {de Palma}, {Desiante}, {Di Mauro}, {Di
  Venere}, {Dom{\'{\i}}nguez}, {Drell}, {Favuzzi}, {Fegan}, {Ferrara}, {Focke},
  {Fortin}, {Franckowiak}, {Fukazawa}, {Funk}, {Furniss}, {Fusco}, {Gargano},
  {Gasparrini}, {Giglietto}, {Giommi}, {Giordano}, {Giroletti}, {Glanzman},
  {Godfrey}, {Grenier}, {Grondin}, {Guillemot}, {Guiriec}, {Harding}, {Hays},
  {Hewitt}, {Hill}, {Horan}, {Iafrate}, {Hartmann}, {Jogler},
  {J{\'o}hannesson}, {Johnson}, {Kamae}, {Kataoka}, {Kn{\"o}dlseder}, {Kuss},
  {La Mura}, {Larsson}, {Latronico}, {Lemoine-Goumard}, {Li}, {Li}, {Longo},
  {Loparco}, {Lott}, {Lovellette}, {Lubrano}, {Madejski}, {Maldera},
  {Manfreda}, {Mayer}, {Mazziotta}, {Michelson}, {Mirabal}, {Mitthumsiri},
  {Mizuno}, {Moiseev}, {Monzani}, {Morselli}, {Moskalenko}, {Murgia}, {Nuss},
  {Ohsugi}, {Omodei}, {Orienti}, {Orlando}, {Ormes}, {Paneque}, {Perkins},
  {Pesce-Rollins}, {Petrosian}, {Piron}, {Pivato}, {Porter}, {Rain{\`o}},
  {Rando}, {Razzano}, {Razzaque}, {Reimer}, {Reimer}, {Reposeur}, {Romani},
  {S{\'a}nchez-Conde}, {Saz Parkinson}, {Schmid}, {Schulz}, {Sgr{\`o}},
  {Siskind}, {Spada}, {Spandre}, {Spinelli}, {Suson}, {Tajima}, {Takahashi},
  {Takahashi}, {Takahashi}, {Thayer}, {Thompson}, {Tibaldo}, {Torres}, {Tosti},
  {Troja}, {Vianello}, {Wood}, {Wood}, {Yassine}, {Zaharijas}, \&
  {Zimmer}}]{2FHL}
---. 2016, \apjs, 222, 5

\bibitem[{{Ahlers} \& {Halzen}(2015)}]{2015RPPh...78l6901A}
{Ahlers}, M., \& {Halzen}, F. 2015, Reports on Progress in Physics, 78, 126901

\bibitem[{Ahlers \& Halzen(2018)}]{Ahlers:2018fkn}
Ahlers, M., \& Halzen, F. 2018, Prog. Part. Nucl. Phys., 102, 73

\bibitem[{{Ahn} {et~al.}(2012){Ahn}, {Alexandroff}, {Allende Prieto},
  {Anderson}, {Anderton}, {Andrews}, {Aubourg}, {Bailey}, {Balbinot}, {Barnes},
  \& et~al.}]{GB_redshift_old}
{Ahn}, C.~P., {Alexandroff}, R., {Allende Prieto}, C., {et~al.} 2012, \apjs,
  203, 21

\bibitem[{{Ajello} {et~al.}(2014){Ajello}, {Romani}, {Gasparrini}, {Shaw},
  {Bolmer}, {Cotter}, {Finke}, {Greiner}, {Healey}, {King}, {Max-Moerbeck},
  {Michelson}, {Potter}, {Rau}, {Readhead}, {Richards}, \&
  {Schady}}]{2014ApJ...780...73A}
{Ajello}, M., {Romani}, R.~W., {Gasparrini}, D., {et~al.} 2014, \apj, 780, 73

\bibitem[{{Ajello} {et~al.}(2017){Ajello}, {Atwood}, {Baldini}, {Ballet},
  {Barbiellini}, {Bastieri}, {Bellazzini}, {Bissaldi}, {Blandford}, {Bloom},
  {Bonino}, {Bregeon}, {Britto}, {Bruel}, {Buehler}, {Buson}, {Cameron},
  {Caputo}, {Caragiulo}, {Caraveo}, {Cavazzuti}, {Cecchi}, {Charles},
  {Chekhtman}, {Cheung}, {Chiaro}, {Ciprini}, {Cohen}, {Costantin}, {Costanza},
  {Cuoco}, {Cutini}, {D'Ammando}, {de Palma}, {Desiante}, {Digel}, {Di Lalla},
  {Di Mauro}, {Di Venere}, {Dom{\'{\i}}nguez}, {Drell}, {Dumora}, {Favuzzi},
  {Fegan}, {Ferrara}, {Fortin}, {Franckowiak}, {Fukazawa}, {Funk}, {Fusco},
  {Gargano}, {Gasparrini}, {Giglietto}, {Giommi}, {Giordano}, {Giroletti},
  {Glanzman}, {Green}, {Grenier}, {Grondin}, {Grove}, {Guillemot}, {Guiriec},
  {Harding}, {Hays}, {Hewitt}, {Horan}, {J{\'o}hannesson}, {Kensei}, {Kuss},
  {La Mura}, {Larsson}, {Latronico}, {Lemoine-Goumard}, {Li}, {Longo},
  {Loparco}, {Lott}, {Lubrano}, {Magill}, {Maldera}, {Manfreda}, {Mazziotta},
  {McEnery}, {Meyer}, {Michelson}, {Mirabal}, {Mitthumsiri}, {Mizuno},
  {Moiseev}, {Monzani}, {Morselli}, {Moskalenko}, {Negro}, {Nuss}, {Ohsugi},
  {Omodei}, {Orienti}, {Orlando}, {Palatiello}, {Paliya}, {Paneque}, {Perkins},
  {Persic}, {Pesce-Rollins}, {Piron}, {Porter}, {Principe}, {Rain{\`o}},
  {Rando}, {Razzano}, {Razzaque}, {Reimer}, {Reimer}, {Reposeur}, {Saz
  Parkinson}, {Sgr{\`o}}, {Simone}, {Siskind}, {Spada}, {Spandre}, {Spinelli},
  {Stawarz}, {Suson}, {Takahashi}, {Tak}, {Thayer}, {Thayer}, {Thompson},
  {Torres}, {Torresi}, {Troja}, {Vianello}, {Wood}, \&
  {Wood}}]{2017ApJS..232...18A}
{Ajello}, M., {Atwood}, W.~B., {Baldini}, L., {et~al.} 2017, \apjs, 232, 18

\bibitem[{{Alard}(2000)}]{alard00}
{Alard}, C. 2000, \aaps, 144, 363

\bibitem[{{Alard} \& {Lupton}(1998)}]{alard98}
{Alard}, C., \& {Lupton}, R.~H. 1998, \apj, 503, 325

\bibitem[{{Albert} {et~al.}(2018){Albert}, {Andr{\'e}}, {Anghinolfi}, {Anton},
  {Ardid}, {Aubert}, {Aublin}, {Avgitas}, {Baret}, {Barrios-Mart{\'{\i}}},
  {Basa}, {Belhorma}, {Bertin}, {Biagi}, {Bormuth}, {Boumaaza}, {Bourret},
  {Bouwhuis}, {Br{\^a}nza{\c s}}, {Bruijn}, {Brunner}, {Busto}, {Capone},
  {Caramete}, {Carr}, {Celli}, {Chabab}, {Cherkaoui El Moursli}, {Chiarusi},
  {Circella}, {Coelho}, {Coleiro}, {Colomer}, {Coniglione}, {Costantini},
  {Coyle}, {Creusot}, {D{\'{\i}}az}, {Deschamps}, {Distefano}, {Di Palma},
  {Domi}, {Don{\`a}}, {Donzaud}, {Dornic}, {Drouhin}, {Eberl}, {El Bojaddaini},
  {El Khayati}, {Els{\"a}sser}, {Enzenh{\"o}fer}, {Ettahiri}, {Fassi}, {Felis},
  {Fermani}, {Ferrara}, {Fusco}, {Gay}, {Glotin}, {Gr{\'e}goire}, {Ruiz},
  {Graf}, {Hallmann}, {van Haren}, {Heijboer}, {Hello}, {Hern{\'a}ndez-Rey},
  {H{\"o}{\ss}l}, {Hofest{\"a}dt}, {Illuminati}, {de Jong}, {Jongen}, {Kadler},
  {Kalekin}, {Katz}, {Khan-Chowdhury}, {Kouchner}, {Kreter}, {Kreykenbohm},
  {Kulikovskiy}, {Lachaud}, {Lahmann}, {Lef{\`e}vre}, {Leonora}, {Lotze},
  {Loucatos}, {Marcelin}, {Margiotta}, {Marinelli}, {Mart{\'{\i}}nez-Mora},
  {Mele}, {Melis}, {Migliozzi}, {Moussa}, {Navas}, {Nezri}, {Nu{\~n}ez},
  {Organokov}, {P{\u a}v{\u a}la{\c s}}, {Pellegrino}, {Piattelli}, {Popa},
  {Pradier}, {Quinn}, {Racca}, {Randazzo}, {Riccobene}, {S{\'a}nchez-Losa},
  {Salda{\~n}a}, {Salvadori}, {Samtleben}, {Sanguineti}, {Sapienza},
  {Sch{\"u}ssler}, {Spurio}, {Stolarczyk}, {Taiuti}, {Tayalati}, {Trovato},
  {Vallage}, {Van Elewyck}, {Versari}, {Vivolo}, {Wilms}, {Zaborov}, {Zornoza},
  {Z{\'u}{\~n}iga}, \& {ANTARES Collaboration}}]{2018ApJ...863L..30A}
{Albert}, A., {Andr{\'e}}, M., {Anghinolfi}, M., {et~al.} 2018, \apjl, 863, L30

\bibitem[{{Ansoldi} {et~al.}(2018){Ansoldi}, {Antonelli}, {Arcaro}, {Baack},
  {Babi{\'c}}, {Banerjee}, {Bangale}, {Barres de Almeida}, {Barrio}, {Becerra
  Gonz{\'a}lez}, {Bednarek}, {Bernardini}, {Berse}, {Berti}, {Besenrieder},
  {Bhattacharyya}, {Bigongiari}, {Biland}, {Blanch}, {Bonnoli}, {Carosi},
  {Ceribella}, {Chatterjee}, {Colak}, {Colin}, {Colombo}, {Contreras},
  {Cortina}, {Covino}, {Cumani}, {D'Elia}, {Da Vela}, {Dazzi}, {De Angelis},
  {De Lotto}, {Delfino}, {Delgado}, {Di Pierro}, {Dom{\'{\i}}nguez}, {Dominis
  Prester}, {Dorner}, {Doro}, {Einecke}, {Elsaesser}, {Fallah Ramazani},
  {Fattorini}, {Fern{\'a}ndez-Barral}, {Ferrara}, {Fidalgo}, {Foffano},
  {Fonseca}, {Font}, {Fruck}, {Gallozzi}, {Garc{\'{\i}}a L{\'o}pez},
  {Garczarczyk}, {Gaug}, {Giammaria}, {Godinovi{\'c}}, {Guberman}, {Hadasch},
  {Hahn}, {Hassan}, {Hayashida}, {Herrera}, {Hoang}, {Hrupec}, {Inoue},
  {Ishio}, {Iwamura}, {Konno}, {Kubo}, {Kushida}, {Lamastra}, {Lelas}, {Leone},
  {Lindfors}, {Lombardi}, {Longo}, {L{\'o}pez}, {Maggio}, {Majumdar},
  {Makariev}, {Maneva}, {Manganaro}, {Mannheim}, {Maraschi}, {Mariotti},
  {Mart{\'{\i}}nez}, {Masuda}, {Mazin}, {Mielke}, {Minev}, {Miranda},
  {Mirzoyan}, {Moralejo}, {Moreno}, {Moretti}, {Neustroev}, {Niedzwiecki},
  {Nievas Rosillo}, {Nigro}, {Nilsson}, {Ninci}, {Nishijima}, {Noda},
  {Nogu{\'e}s}, {Paiano}, {Palacio}, {Paneque}, {Paoletti}, {Paredes},
  {Pedaletti}, {Pe{\~n}il}, {Peresano}, {Persic}, {Pfrang}, {Prada Moroni},
  {Prandini}, {Puljak}, {Garcia}, {Rhode}, {Rib{\'o}}, {Rico}, {Righi},
  {Rugliancich}, {Saha}, {Saito}, {Satalecka}, {Schweizer}, {Sitarek}, {{\v
  S}nidari{\'c}}, {Sobczynska}, {Stamerra}, {Strzys}, {Suri{\'c}}, {Tavecchio},
  {Temnikov}, {Terzi{\'c}}, {Teshima}, {Torres-Alb{\'a}}, {Tsujimoto}, {Vanzo},
  {Vazquez Acosta}, {Vovk}, {Ward}, {Will}, {Zari{\'c}}, \&
  {Cerruti}}]{Ahnen:2018mvi}
{Ansoldi}, S., {Antonelli}, L.~A., {Arcaro}, C., {et~al.} 2018, \apjl, 863, L10

\bibitem[{{Astropy Collaboration} {et~al.}(2013){Astropy Collaboration},
  {Robitaille}, {Tollerud}, {Greenfield}, {Droettboom}, {Bray}, {Aldcroft},
  {Davis}, {Ginsburg}, {Price-Whelan}, {Kerzendorf}, {Conley}, {Crighton},
  {Barbary}, {Muna}, {Ferguson}, {Grollier}, {Parikh}, {Nair}, {Unther},
  {Deil}, {Woillez}, {Conseil}, {Kramer}, {Turner}, {Singer}, {Fox}, {Weaver},
  {Zabalza}, {Edwards}, {Azalee Bostroem}, {Burke}, {Casey}, {Crawford},
  {Dencheva}, {Ely}, {Jenness}, {Labrie}, {Lim}, {Pierfederici}, {Pontzen},
  {Ptak}, {Refsdal}, {Servillat}, \& {Streicher}}]{astropy:2013}
{Astropy Collaboration}, {Robitaille}, T.~P., {Tollerud}, E.~J., {et~al.} 2013,
  \aap, 558, A33

\bibitem[{{Atoyan} \& {Dermer}(2001)}]{atoyan01}
{Atoyan}, A., \& {Dermer}, C.~D. 2001, Physical Review Letters, 87, 221102

\bibitem[{Atoyan \& Dermer(2008)}]{Atoyan:2008uy}
Atoyan, A., \& Dermer, C.~D. 2008, Astrophys. J., 687, L75

\bibitem[{{Atoyan} \& {Dermer}(2003{\natexlab{a}})}]{atoyan03}
{Atoyan}, A.~M., \& {Dermer}, C.~D. 2003{\natexlab{a}}, \apj, 586, 79

\bibitem[{{Atoyan} \& {Dermer}(2003{\natexlab{b}})}]{2003ApJ...586...79A}
---. 2003{\natexlab{b}}, \apj, 586, 79

\bibitem[{{Atwood} {et~al.}(2009){Atwood}, {Abdo}, {Ackermann},
  {et~al.}}]{2009ApJ...697.1071A}
{Atwood}, W.~B., {Abdo}, A.~A., {Ackermann}, M., {et~al.} 2009, \apj, 697, 1071

\bibitem[{Bartos \& Kowalski(2017)}]{10.1088/978-0-7503-1369-8}
Bartos, I., \& Kowalski, M. 2017, Multimessenger Astronomy, 2399-2891 (IOP
  Publishing), doi:10.1088/978-0-7503-1369-8.
\newblock \url{http://dx.doi.org/10.1088/978-0-7503-1369-8}

\bibitem[{{Bednarek} \& {Protheroe}(1999)}]{bednarek99}
{Bednarek}, W., \& {Protheroe}, R.~J. 1999, \mnras, 302, 373

\bibitem[{{Bianchi} {et~al.}(2017){Bianchi}, {Shiao}, \&
  {Thilker}}]{2017ApJS..230...24B}
{Bianchi}, L., {Shiao}, B., \& {Thilker}, D. 2017, \apjs, 230, 24

\bibitem[{{B{\"o}ttcher}(2005)}]{2005ApJ...621..176B}
{B{\"o}ttcher}, M. 2005, \apj, 621, 176

\bibitem[{{B{\"o}ttcher} {et~al.}(2013{\natexlab{a}}){B{\"o}ttcher}, {Reimer},
  {Sweeney}, \& {Prakash}}]{boettcher13}
{B{\"o}ttcher}, M., {Reimer}, A., {Sweeney}, K., \& {Prakash}, A.
  2013{\natexlab{a}}, \apj, 768, 54

\bibitem[{{B{\"o}ttcher} {et~al.}(2013{\natexlab{b}}){B{\"o}ttcher}, {Reimer},
  {Sweeney}, \& {Prakash}}]{2013ApJ...768...54B}
---. 2013{\natexlab{b}}, \apj, 768, 54

\bibitem[{{Cerruti} {et~al.}(2019){Cerruti}, {Zech}, {Boisson}, {Emery},
  {Inoue}, \& {Lenain}}]{Cerruti:2018tmc}
{Cerruti}, M., {Zech}, A., {Boisson}, C., {et~al.} 2019, \mnras, 483, L12

\bibitem[{Cerruti {et~al.}(2015)Cerruti, Zech, Boisson, \&
  Inoue}]{Cerruti:2014iwa}
Cerruti, M., Zech, A., Boisson, C., \& Inoue, S. 2015, Mon. Not. Roy. Astron.
  Soc., 448, 910

\bibitem[{{Dermer} {et~al.}(2012){Dermer}, {Murase}, \& {Takami}}]{dermer12}
{Dermer}, C.~D., {Murase}, K., \& {Takami}, H. 2012, \apj, 755, 147

\bibitem[{Dermer {et~al.}(2012)Dermer, Murase, \& Takami}]{Dermer:2012rg}
Dermer, C.~D., Murase, K., \& Takami, H. 2012, Astrophys. J., 755, 147

\bibitem[{{Dermer} {et~al.}(2009){Dermer}, {Razzaque}, {Finke}, \&
  {Atoyan}}]{dermer09}
{Dermer}, C.~D., {Razzaque}, S., {Finke}, J.~D., \& {Atoyan}, A. 2009, New
  Journal of Physics, 11, 065016

\bibitem[{Dermer {et~al.}(2009)Dermer, Razzaque, Finke, \&
  Atoyan}]{Dermer:2008cy}
Dermer, C.~D., Razzaque, S., Finke, J.~D., \& Atoyan, A. 2009, New J. Phys.,
  11, 065016

\bibitem[{Diltz {et~al.}(2015)Diltz, Boettcher, \& Fossati}]{Diltz:2015kha}
Diltz, C., Boettcher, M., \& Fossati, G. 2015, Astrophys. J., 802, 133

\bibitem[{{Dimitrakoudis} {et~al.}(2012){Dimitrakoudis}, {Mastichiadis},
  {Protheroe}, \& {Reimer}}]{dimitrakoudis12}
{Dimitrakoudis}, S., {Mastichiadis}, A., {Protheroe}, R.~J., \& {Reimer}, A.
  2012, \aap, 546, A120

\bibitem[{{Drake} {et~al.}(2009){Drake}, {Djorgovski}, {Mahabal}, {Beshore},
  {Larson}, {Graham}, {Williams}, {Christensen}, {Catelan}, {Boattini},
  {Gibbs}, {Hill}, \& {Kowalski}}]{2009ApJ...696..870D}
{Drake}, A.~J., {Djorgovski}, S.~G., {Mahabal}, A., {et~al.} 2009, \apj, 696,
  870

\bibitem[{{Evans} {et~al.}(2014){Evans}, {Osborne}, {Beardmore}, {Page},
  {Willingale}, {Mountford}, {Pagani}, {Burrows}, {Kennea}, {Perri},
  {Tagliaferri}, \& {Gehrels}}]{2014ApJS..210....8E}
{Evans}, P.~A., {Osborne}, J.~P., {Beardmore}, A.~P., {et~al.} 2014, \apjs,
  210, 8

\bibitem[{Fujita {et~al.}(2015)Fujita, Kimura, \& Murase}]{Fujita:2015xva}
Fujita, Y., Kimura, S.~S., \& Murase, K. 2015, Phys. Rev., D92, 023001

\bibitem[{Gao {et~al.}(2019)Gao, Fedynitch, Winter, \& Pohl}]{Gao:2018mnu}
Gao, S., Fedynitch, A., Winter, W., \& Pohl, M. 2019, Nat. Astron., 3, 88

\bibitem[{{Gregory} {et~al.}(1996){Gregory}, {Scott}, {Douglas}, \&
  {Condon}}]{1996ApJS..103..427G}
{Gregory}, P.~C., {Scott}, W.~K., {Douglas}, K., \& {Condon}, J.~J. 1996,
  \apjs, 103, 427

\bibitem[{{Halzen}(2013)}]{halzen13}
{Halzen}, F. 2013, Astroparticle Physics, 43, 155

\bibitem[{{Helfand} {et~al.}(2015){Helfand}, {White}, \&
  {Becker}}]{2015ApJ...801...26H}
{Helfand}, D.~J., {White}, R.~L., \& {Becker}, R.~H. 2015, \apj, 801, 26

\bibitem[{{Henden} {et~al.}(2015){Henden}, {Levine}, {Terrell}, \&
  {Welch}}]{henden15}
{Henden}, A.~A., {Levine}, S., {Terrell}, D., \& {Welch}, D.~L. 2015, in
  American Astronomical Society Meeting Abstracts, Vol. 225, American
  Astronomical Society Meeting Abstracts \#225, 336.16

\bibitem[{{IceCube Collaboration} {et~al.}(2017){IceCube Collaboration},
  {Aartsen}, {Ackermann}, {Adams}, {Aguilar}, {Ahlers}, {Ahrens}, {Samarai},
  {Altmann}, {Andeen}, \& et~al.}]{2017arXiv171001191I}
{IceCube Collaboration}, {Aartsen}, M.~G., {Ackermann}, M., {et~al.} 2017,
  ArXiv e-prints, arXiv:1710.01191

\bibitem[{Kachelriess {et~al.}(2009)Kachelriess, Ostapchenko, \&
  Tomas}]{Kachelriess:2008qx}
Kachelriess, M., Ostapchenko, S., \& Tomas, R. 2009, New J. Phys., 11, 065017

\bibitem[{{Kadler} {et~al.}(2016){Kadler}, {Krau{\ss}}, {Mannheim}, {Ojha},
  {M{\"u}ller}, {Schulz}, {Anton}, {Baumgartner}, {Beuchert}, {Buson},
  {Carpenter}, {Eberl}, {Edwards}, {Eisenacher Glawion}, {Els{\"a}sser},
  {Gehrels}, {Gr{\"a}fe}, {Gulyaev}, {Hase}, {Horiuchi}, {James}, {Kappes},
  {Kappes}, {Katz}, {Kreikenbohm}, {Kreter}, {Kreykenbohm}, {Langejahn},
  {Leiter}, {Litzinger}, {Longo}, {Lovell}, {McEnery}, {Natusch}, {Phillips},
  {Pl{\"o}tz}, {Quick}, {Ros}, {Stecker}, {Steinbring}, {Stevens}, {Thompson},
  {Tr{\"u}stedt}, {Tzioumis}, {Weston}, {Wilms}, \&
  {Zensus}}]{2016NatPh..12..807K}
{Kadler}, M., {Krau{\ss}}, F., {Mannheim}, K., {et~al.} 2016, Nature Physics,
  12, 807

\bibitem[{{Keivani} {et~al.}(2018){Keivani}, {Murase}, {Petropoulou}, {Fox},
  {Cenko}, {Chaty}, {Coleiro}, {DeLaunay}, {Dimitrakoudis}, {Evans}, {Kennea},
  {Marshall}, {Mastichiadis}, {Osborne}, {Santander}, {Tohuvavohu}, \&
  {Turley}}]{Keivani:2018rnh}
{Keivani}, A., {Murase}, K., {Petropoulou}, M., {et~al.} 2018, \apj, 864, 84

\bibitem[{{Kochanek} {et~al.}(2017){Kochanek}, {Shappee}, {Stanek}, {Holoien},
  {Thompson}, {Prieto}, {Dong}, {Shields}, {Will}, {Britt}, {Perzanowski}, \&
  {Pojma{\'n}ski}}]{2017PASP..129j4502K}
{Kochanek}, C.~S., {Shappee}, B.~J., {Stanek}, K.~Z., {et~al.} 2017, \pasp,
  129, 104502

\bibitem[{{Krau{\ss}} {et~al.}(2018){Krau{\ss}}, {Deoskar}, {Baxter}, {Kadler},
  {Kreter}, {Langejahn}, {Mannheim}, {Polko}, {Wang}, \&
  {Wilms}}]{2018A&A...620A.174K}
{Krau{\ss}}, F., {Deoskar}, K., {Baxter}, C., {et~al.} 2018, \aap, 620, A174

\bibitem[{{Laher} {et~al.}(2014){Laher}, {Surace}, {Grillmair}, {Ofek},
  {Levitan}, {Sesar}, {van Eyken}, {Law}, {Helou}, {Hamam}, {Masci},
  {Mattingly}, {Jackson}, {Hacopeans}, {Mi}, {Groom}, {Teplitz}, {Desai},
  {Hale}, {Smith}, {Walters}, {Quimby}, {Kasliwal}, {Horesh}, {Bellm},
  {Barlow}, {Waszczak}, {Prince}, \& {Kulkarni}}]{2014PASP..126..674L}
{Laher}, R.~R., {Surace}, J., {Grillmair}, C.~J., {et~al.} 2014, \pasp, 126,
  674

\bibitem[{{Lambert} \& {Gontier}(2009)}]{2009A&A...493..317L}
{Lambert}, S.~B., \& {Gontier}, A.~M. 2009, \aap, 493, 317

\bibitem[{Liang {et~al.}(2018)Liang, He, Liao, Xin, Yuan, \&
  Fan}]{Liang:2018siw}
Liang, Y.-F., He, H.-N., Liao, N.-H., {et~al.} 2018, arXiv:1807.05057

\bibitem[{{Liu} {et~al.}(2018){Liu}, {Wang}, {Xue}, {Taylor}, {Wang}, {Li}, \&
  {Yan}}]{2018arXiv180705113L}
{Liu}, R.-Y., {Wang}, K., {Xue}, R., {et~al.} 2018, ArXiv e-prints,
  arXiv:1807.05113

\bibitem[{{Lott} {et~al.}(2012){Lott}, {Escande}, {Larsson}, \&
  {Ballet}}]{2012A&A...544A...6L}
{Lott}, B., {Escande}, L., {Larsson}, S., \& {Ballet}, J. 2012, \aap, 544, A6

\bibitem[{{Lucarelli} {et~al.}(2019){Lucarelli}, {Tavani}, {Piano},
  {Bulgarelli}, {Donnarumma}, {Verrecchia}, {Pittori}, {Antonelli}, {Argan},
  {Barbiellini}, {Caraveo}, {Cardillo}, {Cattaneo}, {Chen}, {Colafrancesco},
  {Costa}, {Del Monte}, {Di Cocco}, {Ferrari}, {Fioretti}, {Galli}, {Giommi},
  {Giuliani}, {Lipari}, {Longo}, {Mereghetti}, {Morselli}, {Paoletti},
  {Parmiggiani}, {Pellizzoni}, {Picozza}, {Pilia}, {Rappoldi}, {Trois}, {Ursi},
  {Vercellone}, {Vittorini}, \& {The AGILE Team}}]{2019ApJ...870..136L}
{Lucarelli}, F., {Tavani}, M., {Piano}, G., {et~al.} 2019, \apj, 870, 136

\bibitem[{{Mannheim}(1993)}]{1993A&A...269...67M}
{Mannheim}, K. 1993, \aap, 269, 67

\bibitem[{{Mannheim}(1995)}]{1995APh.....3..295M}
---. 1995, Astroparticle Physics, 3, 295

\bibitem[{{Mannheim} \& {Biermann}(1989)}]{mannheim89}
{Mannheim}, K., \& {Biermann}, P.~L. 1989, \aap, 221, 211

\bibitem[{{Mannheim} {et~al.}(2001){Mannheim}, {Protheroe}, \&
  {Rachen}}]{mannheim01}
{Mannheim}, K., {Protheroe}, R.~J., \& {Rachen}, J.~P. 2001, \prd, 63, 023003

\bibitem[{{Mannheim} {et~al.}(1992){Mannheim}, {Stanev}, \&
  {Biermann}}]{mannheim92}
{Mannheim}, K., {Stanev}, T., \& {Biermann}, P.~L. 1992, \aap, 260, L1

\bibitem[{{Masci} {et~al.}(2017){Masci}, {Laher}, {Rebbapragada}, {Doran},
  {Miller}, {Bellm}, {Kasliwal}, {Ofek}, {Surace}, {Shupe}, {Grillmair},
  {Jackson}, {Barlow}, {Yan}, {Cao}, {Cenko}, {Storrie-Lombardi}, {Helou},
  {Prince}, \& {Kulkarni}}]{2017PASP..129a4002M}
{Masci}, F.~J., {Laher}, R.~R., {Rebbapragada}, U.~D., {et~al.} 2017, \pasp,
  129, 014002

\bibitem[{Maselli {et~al.}(2015)Maselli, Massaro, D'Abrusco, Cusumano,
  La~Parola, Segreto, \& Tosti}]{Maselli:2015oha}
Maselli, A., Massaro, F., D'Abrusco, R., {et~al.} 2015, Astrophys. Space Sci.,
  357, 141

\bibitem[{Massaro {et~al.}(2004)Massaro, Perri, Giommi, \&
  Nesci}]{Massaro:2003sx}
Massaro, E., Perri, M., Giommi, P., \& Nesci, R. 2004, Astron. Astrophys., 413,
  489

\bibitem[{{Mastichiadis}(1996)}]{mastichiadis96}
{Mastichiadis}, A. 1996, \ssr, 75, 317

\bibitem[{{Mattox} {et~al.}(1996){Mattox}, {Bertsch}, {Chiang}, {Dingus},
  {Digel}, {Esposito}, {Fierro}, {Hartman}, {Hunter}, {Kanbach}, {Kniffen},
  {Lin}, {Macomb}, {Mayer-Hasselwander}, {Michelson}, {von Montigny},
  {Mukherjee}, {Nolan}, {Ramanamurthy}, {Schneid}, {Sreekumar}, {Thompson}, \&
  {Willis}}]{mattox96}
{Mattox}, J.~R., {Bertsch}, D.~L., {Chiang}, J., {et~al.} 1996, \apj, 461, 396

\bibitem[{{M{\"u}cke} \& {Protheroe}(2001)}]{mucke01}
{M{\"u}cke}, A., \& {Protheroe}, R.~J. 2001, Astroparticle Physics, 15, 121

\bibitem[{{M{\"u}cke} {et~al.}(2003){M{\"u}cke}, {Protheroe}, {Engel},
  {Rachen}, \& {Stanev}}]{mucke03}
{M{\"u}cke}, A., {Protheroe}, R.~J., {Engel}, R., {Rachen}, J.~P., \& {Stanev},
  T. 2003, Astroparticle Physics, 18, 593

\bibitem[{{Murase} {et~al.}(2016){Murase}, {Guetta}, \&
  {Ahlers}}]{2016PhRvL.116g1101M}
{Murase}, K., {Guetta}, D., \& {Ahlers}, M. 2016, Physical Review Letters, 116,
  071101

\bibitem[{{Murase} {et~al.}(2018){Murase}, {Oikonomou}, \&
  {Petropoulou}}]{Murase:2018iyl}
{Murase}, K., {Oikonomou}, F., \& {Petropoulou}, M. 2018, \apj, 865, 124

\bibitem[{{Padovani} {et~al.}(2018){Padovani}, {Giommi}, {Resconi}, {Glauch},
  {Arsioli}, {Sahakyan}, \& {Huber}}]{2018arXiv180704461P}
{Padovani}, P., {Giommi}, P., {Resconi}, E., {et~al.} 2018, \mnras, 480, 192

\bibitem[{{Padovani} \& {Resconi}(2014)}]{2014MNRAS.443..474P}
{Padovani}, P., \& {Resconi}, E. 2014, \mnras, 443, 474

\bibitem[{{Padovani} {et~al.}(2016){Padovani}, {Resconi}, {Giommi}, {Arsioli},
  \& {Chang}}]{2016MNRAS.457.3582P}
{Padovani}, P., {Resconi}, E., {Giommi}, P., {Arsioli}, B., \& {Chang}, Y.~L.
  2016, \mnras, 457, 3582

\bibitem[{{Paiano} {et~al.}(2018){Paiano}, {Falomo}, {Treves}, \&
  {Scarpa}}]{2018ApJ...854L..32P}
{Paiano}, S., {Falomo}, R., {Treves}, A., \& {Scarpa}, R. 2018, \apjl, 854, L32

\bibitem[{Petropoulou \& Dimitrakoudis(2015)}]{Petropoulou:2015swa}
Petropoulou, M., \& Dimitrakoudis, S. 2015, Mon. Not. Roy. Astron. Soc., 452,
  1303

\bibitem[{{Price-Whelan} {et~al.}(2018){Price-Whelan}, {Sip{\H{o}}cz},
  {G{\"u}nther}, {Lim}, {Crawford}, {Conseil}, {Shupe}, {Craig}, {Dencheva},
  {Ginsburg}, {VanderPlas}, {Bradley}, {P{\'e}rez-Su{\'a}rez}, {de Val-Borro},
  {Paper Contributors}, {Aldcroft}, {Cruz}, {Robitaille}, {Tollerud},
  {Coordination Committee}, {Ardelean}, {Babej}, {Bach}, {Bachetti}, {Bakanov},
  {Bamford}, {Barentsen}, {Barmby}, {Baumbach}, {Berry}, {Biscani}, {Boquien},
  {Bostroem}, {Bouma}, {Brammer}, {Bray}, {Breytenbach}, {Buddelmeijer},
  {Burke}, {Calderone}, {Cano Rodr{\'\i}guez}, {Cara}, {Cardoso}, {Cheedella},
  {Copin}, {Corrales}, {Crichton}, {D{\textquoteright}Avella}, {Deil},
  {Depagne}, {Dietrich}, {Donath}, {Droettboom}, {Earl}, {Erben}, {Fabbro},
  {Ferreira}, {Finethy}, {Fox}, {Garrison}, {Gibbons}, {Goldstein}, {Gommers},
  {Greco}, {Greenfield}, {Groener}, {Grollier}, {Hagen}, {Hirst}, {Homeier},
  {Horton}, {Hosseinzadeh}, {Hu}, {Hunkeler}, {Ivezi{\'c}}, {Jain}, {Jenness},
  {Kanarek}, {Kendrew}, {Kern}, {Kerzendorf}, {Khvalko}, {King}, {Kirkby},
  {Kulkarni}, {Kumar}, {Lee}, {Lenz}, {Littlefair}, {Ma}, {Macleod},
  {Mastropietro}, {McCully}, {Montagnac}, {Morris}, {Mueller}, {Mumford},
  {Muna}, {Murphy}, {Nelson}, {Nguyen}, {Ninan}, {N{\"o}the}, {Ogaz}, {Oh},
  {Parejko}, {Parley}, {Pascual}, {Patil}, {Patil}, {Plunkett}, {Prochaska},
  {Rastogi}, {Reddy Janga}, {Sabater}, {Sakurikar}, {Seifert}, {Sherbert},
  {Sherwood-Taylor}, {Shih}, {Sick}, {Silbiger}, {Singanamalla}, {Singer},
  {Sladen}, {Sooley}, {Sornarajah}, {Streicher}, {Teuben}, {Thomas},
  {Tremblay}, {Turner}, {Terr{\'o}n}, {van Kerkwijk}, {de la Vega}, {Watkins},
  {Weaver}, {Whitmore}, {Woillez}, {Zabalza}, \& {Contributors}}]{astropy:2018}
{Price-Whelan}, A.~M., {Sip{\H{o}}cz}, B.~M., {G{\"u}nther}, H.~M., {et~al.}
  2018, \aj, 156, 123

\bibitem[{{Protheroe}(1999)}]{protheroe99}
{Protheroe}, R.~J. 1999, Nuclear Physics B Proceedings Supplements, 77, 465

\bibitem[{{Protheroe} {et~al.}(2003){Protheroe}, {Donea}, \&
  {Reimer}}]{protheroe03}
{Protheroe}, R.~J., {Donea}, A.-C., \& {Reimer}, A. 2003, Astroparticle
  Physics, 19, 559

\bibitem[{{Protheroe} \& {Szabo}(1992)}]{protheroe92}
{Protheroe}, R.~J., \& {Szabo}, A.~P. 1992, Physical Review Letters, 69, 2885

\bibitem[{Reimer {et~al.}(2018)Reimer, Boettcher, \& Buson}]{Reimer:2018vvw}
Reimer, A., Boettcher, M., \& Buson, S. 2018, arXiv:1812.05654

\bibitem[{{Reimer} {et~al.}(2004){Reimer}, {Protheroe}, \& {Donea}}]{reimer04}
{Reimer}, A., {Protheroe}, R.~J., \& {Donea}, A.-C. 2004, \aap, 419, 89

\bibitem[{{Richards} {et~al.}(2009){Richards}, {Myers}, {Gray}, {Riegel},
  {Nichol}, {Brunner}, {Szalay}, {Schneider}, \&
  {Anderson}}]{2009ApJS..180...67R}
{Richards}, G.~T., {Myers}, A.~D., {Gray}, A.~G., {et~al.} 2009, \apjs, 180, 67

\bibitem[{Rodrigues {et~al.}(2018)Rodrigues, Gao, Fedynitch, Palladino, \&
  Winter}]{Rodrigues:2018tku}
Rodrigues, X., Gao, S., Fedynitch, A., Palladino, A., \& Winter, W. 2018,
  arXiv:1812.05939

\bibitem[{{Rosen} {et~al.}(2016){Rosen}, {Webb}, {Watson}, {Ballet}, {Barret},
  {Braito}, {Carrera}, {Ceballos}, {Coriat}, {Della Ceca}, {Denkinson},
  {Esquej}, {Farrell}, {Freyberg}, {Gris{\'e}}, {Guillout}, {Heil},
  {Koliopanos}, {Law-Green}, {Lamer}, {Lin}, {Martino}, {Michel}, {Motch},
  {Nebot Gomez-Moran}, {Page}, {Page}, {Page}, {Pakull}, {Pye}, {Read},
  {Rodriguez}, {Sakano}, {Saxton}, {Schwope}, {Scott}, {Sturm}, {Traulsen},
  {Yershov}, \& {Zolotukhin}}]{2016A&A...590A...1R}
{Rosen}, S.~R., {Webb}, N.~A., {Watson}, M.~G., {et~al.} 2016, \aap, 590, A1

\bibitem[{{Scargle} {et~al.}(2013){Scargle}, {Norris}, {Jackson}, \&
  {Chiang}}]{2013ApJ...764..167S}
{Scargle}, J.~D., {Norris}, J.~P., {Jackson}, B., \& {Chiang}, J. 2013, \apj,
  764, 167

\bibitem[{{Shappee} {et~al.}(2014){Shappee}, {Prieto}, {Grupe}, {Kochanek},
  {Stanek}, {De Rosa}, {Mathur}, {Zu}, {Peterson}, {Pogge}, {Komossa}, {Im},
  {Jencson}, {Holoien}, {Basu}, {Beacom}, {Szczygie{\l}}, {Brimacombe},
  {Adams}, {Campillay}, {Choi}, {Contreras}, {Dietrich}, {Dubberley},
  {Elphick}, {Foale}, {Giustini}, {Gonzalez}, {Hawkins}, {Howell}, {Hsiao},
  {Koss}, {Leighly}, {Morrell}, {Mudd}, {Mullins}, {Nugent}, {Parrent},
  {Phillips}, {Pojmanski}, {Rosing}, {Ross}, {Sand}, {Terndrup}, {Valenti},
  {Walker}, \& {Yoon}}]{2014ApJ...788...48S}
{Shappee}, B.~J., {Prieto}, J.~L., {Grupe}, D., {et~al.} 2014, \apj, 788, 48

\bibitem[{{Snellen} {et~al.}(2002){Snellen}, {McMahon}, {Hook}, \&
  {Browne}}]{2002MNRAS.329..700S}
{Snellen}, I.~A.~G., {McMahon}, R.~G., {Hook}, I.~M., \& {Browne}, I.~W.~A.
  2002, \mnras, 329, 700

\bibitem[{{Stecker} {et~al.}(1991){Stecker}, {Done}, {Salamon}, \&
  {Sommers}}]{1991PhRvL..66.2697S}
{Stecker}, F.~W., {Done}, C., {Salamon}, M.~H., \& {Sommers}, P. 1991, Physical
  Review Letters, 66, 2697

\bibitem[{Strotjohann {et~al.}(2018)Strotjohann, Kowalski, \&
  Franckowiak}]{Strotjohann:2018ufz}
Strotjohann, N.~L., Kowalski, M., \& Franckowiak, A. 2018, arXiv:1809.06865

\bibitem[{{Szabo} \& {Protheroe}(1994)}]{szabo94}
{Szabo}, A.~P., \& {Protheroe}, R.~J. 1994, Astroparticle Physics, 2, 375

\bibitem[{{Tanaka} {et~al.}(2017){Tanaka}, {Buson}, \& {Kocevski}}]{Tanaka17}
{Tanaka}, Y.~T., {Buson}, S., \& {Kocevski}, D. 2017, The Astronomer's
  Telegram, 10791

\bibitem[{{Weidinger} \& {Spanier}(2015)}]{2015A&A...573A...7W}
{Weidinger}, M., \& {Spanier}, F. 2015, \aap, 573, A7

\bibitem[{{Weinstein} {et~al.}(2004){Weinstein}, {Richards}, {Schneider},
  {Younger}, {Strauss}, {Hall}, {Budav{\'a}ri}, {Gunn}, {York}, \&
  {Brinkmann}}]{2004ApJS..155..243W}
{Weinstein}, M.~A., {Richards}, G.~T., {Schneider}, D.~P., {et~al.} 2004,
  \apjs, 155, 243

\bibitem[{Wilks(1938)}]{wilks1938}
Wilks, S.~S. 1938, Ann. Math. Statist., 9, 60.
\newblock \url{https://doi.org/10.1214/aoms/1177732360}

\bibitem[{Wood {et~al.}(2018)Wood, Caputo, Charles, Di~Mauro, Magill, \&
  Perkins}]{Wood:2017yyb}
Wood, M., Caputo, R., Charles, E., {et~al.} 2018, PoS, ICRC2017, 824,
  [35,824(2017)]

\bibitem[{{Wright} {et~al.}(2010){Wright}, {Eisenhardt}, {Mainzer}, {Ressler},
  {Cutri}, {Jarrett}, {Kirkpatrick}, {Padgett}, {McMillan}, {Skrutskie},
  {Stanford}, {Cohen}, {Walker}, {Mather}, {Leisawitz}, {Gautier}, {McLean},
  {Benford}, {Lonsdale}, {Blain}, {Mendez}, {Irace}, {Duval}, {Liu}, {Royer},
  {Heinrichsen}, {Howard}, {Shannon}, {Kendall}, {Walsh}, {Larsen}, {Cardon},
  {Schick}, {Schwalm}, {Abid}, {Fabinsky}, {Naes}, \&
  {Tsai}}]{2010AJ....140.1868W}
{Wright}, E.~L., {Eisenhardt}, P. R.~M., {Mainzer}, A.~K., {et~al.} 2010, \aj,
  140, 1868

\bibitem[{{Zhang} {et~al.}(2018){Zhang}, {Fang}, \& {Li}}]{2018arXiv180711069Z}
{Zhang}, H., {Fang}, K., \& {Li}, H. 2018, ArXiv e-prints, arXiv:1807.11069

\end{thebibliography}

\end{document}